\documentclass[preprint, 5p, twocolumn]{elsarticle}

\usepackage{amssymb}
\usepackage{amsmath}

\usepackage[bookmarks=false,colorlinks]{hyperref}
\hypersetup{
    hyperfootnotes=false,
    pdftitle={Photodiamond},    
    pdfauthor={Antonio Cammarata},     
    pdfkeywords={photovoltaics} {diamond} {ab initio} {phonon} {anharmonic} {absorption} {electron-phonon} {coupling} {device},
    colorlinks   = true,
    citecolor=RoyalBlue,
    urlcolor=RoyalBlue,
    linkcolor=RoyalBlue,
    linktocpage=true,
    linktoc=all
}
\usepackage[dvipsnames]{xcolor}
\usepackage{multirow}
\usepackage{graphicx}
\usepackage{pdfpages}

\usepackage{soul}

\journal{Nano Energy}

\begin{document}

\begin{frontmatter}


\title{From Defects to Devices: Design Guidelines for High-Performance Diamond-Based Solar Cells and Single-Dopant Diodes}

\cortext[cor1]{Corresponding author}

\author[label1]{Mat\'{u}\v{s} Kaintz\corref{cor1}}
\ead{kaintmat@fel.cvut.cz}

\author[label1]{Antonio Cammarata\corref{cor1}}
\ead{cammaant@fel.cvut.cz}

\affiliation[label1]{organization={Department of Control Engineering, Faculty of Electrical Engineering, Czech Technical University in Prague},
            addressline={Technicka 2}, 
            city={Prague 6},
            postcode={16627}, 
            country={Czech Republic}}

\begin{abstract}
This work establishes key technological guidelines for designing diamond-based optoelectronic devices, derived from a first-principles investigation of two architectures:
a PIN junction with a boron-vacancy-boron (BVB) intermediate-band absorber, and a PN junction based on phosphorus-vacancy (PV) defects.
For the PIN solar cell, practical design principles include:
\textit{i}) aligning incident light in the \textit{xz}-plane to exploit anisotropic absorption;
\textit{ii}) using graded junctions to mitigate tunnelling losses at abrupt interfaces;
\textit{iii}) targeting an absorber thickness of $\sim$500 nm to balance absorption and carrier extraction; and
\textit{iv}) leveraging the high transparency of both contact layers for bifacial device configurations.
For the PN diode, the PV-doped diamond operates via impurity-band conduction, making it suitable for degenerate p-type applications such as tunnel diodes or asymmetric junctions, while its temperature-dependent Seebeck anisotropy and sign-reversal offer opportunities for thermal management applications.
When paired with phosphorus-doped n-type regions, these defects enable single-dopant junctions that significantly simplify device manufacturing.
Using density functional theory with GW corrections, Bethe-Salpeter equation calculations and carrier transport modelling coupled to device electrostatics via a Poisson solver, we show that the BVB defect introduces intermediate bands without degrading diamond’s high carrier mobility or thermal conductivity, while PV-doping provides high conductivity at room temperature through impurity-band transport.
Overall, both defect-engineered systems preserve diamond's superior transport and thermal properties even after doping, offering viable pathways for high-performance diamond optoelectronics.
These guidelines provide a practical foundation for fabricating efficient diamond-based photovoltaic and diode devices.
\end{abstract}




\end{frontmatter}



\section{Introduction}
\label{sec:introduction}

Owing to its ultra-wide band gap and exceptional transport properties, diamond occupies a unique position among semiconductor materials.
Its combination of high carrier mobility, extreme thermal conductivity, large breakdown field, and chemical and radiation stability enables device operation under conditions that challenge conventional semiconductors \cite{WORT200822, ISBERG2004320, PEREZ2020108154, PANIZZA2005191, 10.1063/1.5034413}.
These properties enable a wide range of applications, including high-frequency field-effect and bipolar transistors \cite{ALEKSOV2003391, KATO201341}, high-power radio-frequency amplification \cite{GURBUZ20051055}, Schottky and PIN diodes \cite{7520776, JHA2021108154, pssa201300051}, electronic switches \cite{WORT200822,PEREZ2020108154}, electrochemical electrodes \cite{PANIZZA2005191}, and radiation detectors \cite{10.1063/1.5034413}.
Beyond electronic and sensing applications, these same properties also make diamond an attractive platform for energy conversion technologies.
In particular, its wide band gap and exceptional transport properties suggest that it could serve as a suitable host for intermediate-band (IB) solar cells, a concept proposed to overcome the Shockley-Queisser limit of conventional single band gap devices \cite{1961JAP....32..510S}.
In single band gap solar cells, a significant portion of the solar spectrum is either not absorbed due to photon energies being below the bandgap or is inefficiently utilised due to carrier thermalisation for photon energies above the bandgap. 
Introducing intermediate bands within the band gap enables absorption of sub-bandgap photons by providing stepping-stone transitions between the valence and conduction bands.
In addition, the intermediate-band concept enables the use of wide band gap semiconductors, thereby improving the utilisation of high-energy photons \cite{PhysRevLett.78.5014, 10.1063/1.1492016, https://doi.org/10.1002/pip.3351, 6729038, Rasukkannu2017}. 
The wide band gap of diamond (5.47 eV) is particularly attractive in this context, as it can host multiple intermediate bands, potentially enabling a large number of optical transitions and high theoretical conversion efficiencies.

To this aim, the first objective of this work is to explore the potential of deep-level impurities in diamond for applications in IB photovoltaics. 
Specifically, we propose a PIN device architecture in which the intrinsic (I) region consists of diamond containing intermediate bands. 
Achieving the desired electronic and optical properties requires controlled introduction of impurities into the diamond lattice to define all three regions of the device.
The most promising candidates are selected based on our previous study of substitutional dopants and defect complexes in diamond \cite{PhysRevApplied.23.054029}.
The P- and N-region are based on boron- and phosphorus-doped diamond, respectively, as both exhibit low activation energies and can be incorporated at high concentrations \cite{PMID:24738731, GROTJOHN2014129}.
The I-region is based on the boron-vacancy-boron (BVB) defect, which introduces multiple intermediate bands suitably distributed within the band gap \cite{PhysRevB.73.085204,PhysRevApplied.23.054029}.
Although direct experimental realisation of BVB complexes remains limited, their formation can be understood based on established defect physics.
Boron-vacancy (BV) complexes are well studied in diamond \cite{PhysRevB.105.165201,MURUGANATHAN2021108341}, while boron-boron (B$_2$) pairs are known to form at high doping concentrations \cite{Ashcheulov2013}.
Furthermore, previous studies indicate that BVB configurations are energetically more favourable than isolated BV complexes \cite{PhysRevB.73.085204, PhysRevApplied.23.054029}.
Combined with the well-established mechanism of vacancy capture by impurity centres under typical growth and annealing conditions (e.g. in the case of nitrogen-related defects \cite{Ashfold2020}), this suggests that BVB complexes can form in diamond under appropriate conditions.

The second objective of this work is to assess the feasibility of a PN junction based on a single dopant species. 
This approach could simplify semiconductor fabrication by eliminating the need for multiple dopants and reducing associated handling and contamination risks.
In this design, phosphorus is used to form the N-region, while the P-region is realised using phosphorus-vacancy (PV) complexes.
This enables a uniform doping strategy, where the P-region is created locally via vacancy engineering.
The PV complex is particularly attractive, as it exhibits p-type behaviour, and diamond doped with vacancy-related defect complexes of comparable size is commonly synthesised \cite{XVcenters, MAKINO2021108248}.
Furthermore, previous studies suggest that the PV complex has a lower formation energy than substitutional phosphorus \cite{PhysRevB.72.035214, PhysRevApplied.23.054029}.

To model the proposed devices, we employ density functional theory (DFT) with GW correction to accurately describe the electronic structure.
Optical properties are evaluated using the Bethe-Salpeter equation, while carrier transport is analysed via electron-phonon coupling calculations.
The resulting first-principles data are then used as input for a Poisson solver to obtain the device-level electrostatics.
Our approach provides first-principles-based framework for assessing the interplay between electronic structure, optical properties, transport, and device-level electrostatics, and establishes physically grounded design guidelines for diamond-based photovoltaic and diode architectures.

\section{Computational details and methodology}
\label{sec:comp_details}

All calculations are performed by means of the \textsc{abinit} package \cite{Verstraete2025, Gonze2020, Gonze2005, Brunin2020b}, within DFT, GW, Bethe-Salpeter equation (BSE) and electron-phonon coupling formalisms.
Following previous studies \cite{CAMMARATA2022109237, PhysRevApplied.23.054029}, we adopt the GGA-WC energy functional \cite{PhysRevB.73.235116} for all of our DFT simulations,  which provides the best agreement of lattice parameters and band gap in pristine diamond with experiments.
The plane-wave energy cutoff is set to 816 eV.
The Brillouin zone is sampled using a $\Gamma$-centred Monkhorst-Pack scheme \cite{PhysRevB.13.5188}, with the k-point density scaled according to the supercell size, i.e. $15\times15\times15$ for pristine diamond and $3\times3\times3$ for $5\times5\times5$ supercells.
Furthermore, the Fermi-Dirac smearing is applied with a smearing temperature of 300 K.
Self-consistent field (SCF) convergence is achieved when the total energy difference between successive iterations falls below $10^{-11}$ eV.
To obtain ground-state structural models, we perform geometry optimisation until the largest component of the forces acting on the atoms is below $2.5\times 10^{-5}$ eV/\AA{}.
With these settings, the optimised pristine-diamond structure exhibits the cubic space group symmetry $Fd\bar{3}m$ (group number 227) with a lattice parameter of 3.560 \AA{}, in good agreement with the experimental value of 3.567 \AA{} \cite{Bindzus:pc5033}.
The calculated band gap of 4.09 eV is consistent with previous DFT studies \cite{article}, although it underestimates the experimental value of 5.47 eV \cite{WORT200822, ISBERG2004320}, as expected from pure DFT calculation \cite{doi:10.1021/acs.inorgchem.9b01785}.

To correct the band gap underestimation and obtain accurate impurity level positions, we perform one-shot GW (often denoted as G$_0$W$_0$) calculations \cite{PhysRev.139.A796} using the Godby-Needs plasmon-pole model \cite{Godby1989}.
Our convergence tests indicate that at least five times the number of occupied bands are required for reliable results in large supercells, with a dielectric matrix plane-wave cutoff of 190 eV.
For pristine diamond, the GW calculation yields a band gap of 5.41 eV, in very good agreement with experimental value.

We compute optical absorption spectra including excitonic effects by solving the Bethe-Salpeter equation (BSE) on top of the GW-corrected eigenvalues within the Tamm-Dancoff approximation.
A dielectric matrix plane-wave cutoff of 109 eV is found sufficient for BSE calculations.
To obtain absorption spectra of boron- and phosphorus-doped diamond we utilise the \texttt{optic} utility within the \textsc{abinit} package, which relies on the independent particle approximation (IPA) \cite{Sharma_2004}, with GW-corrected eigenvalues.
As these systems are degenerately doped and show metallic-like properties we expect the excitonic effects to be significantly screened due to the presence of free carriers.
Additionally, the \texttt{optic} utility is used to qualitatively investigate excited-state absorption in BVB-doped systems, in this case without GW corrections to account for changes in the electronic structure induced by excited occupations.
Optical properties are calculated with the $6\times6\times6$ Monkhorst-Pack mesh for all supercells.

Electron-phonon coupling and corresponding transport properties are computed using the Density Functional Perturbation Theory framework on a coarse $2\times2\times2$ q-point grid, followed by Fourier interpolation of the scattering potentials to denser meshes.
Brillouin-zone integrations are performed using a double-grid scheme, where electron-phonon matrix elements are evaluated on a $6\times6\times6$ grid and combined with electronic energies and phonon frequencies obtained on a finer $12\times12\times12$ grid \cite{Brunin2020b}.
For the BVB system, transport properties are evaluated using a $4\times4\times4$ supercell, corresponding to a defect concentration of 0.78\%, due to numerical challenges associated with electron-phonon calculations in the larger 249-atom supercell.
Nevertheless, the reported carrier mobilities in \autoref{sec:transport} are expected to represent a lower bound for the device-relevant concentration of 0.4$\%$, as reduced defect concentrations lead to weaker perturbations of the host band structure and, consequently, higher carrier mobilities.
All of the reported transport properties are evaluated within the momentum relaxation time approximation (MRTA) \cite{Ponce_2020}.
Furthermore, we calculate the lattice thermal conductivity using the linearized Boltzmann transport equation (LBTE) as implemented in the \texttt{phono3py} code \cite{phono3py, phonopy-phono3py-JPCM, PhysRevLett.110.265506}. 
Due to the computational cost of third-order force constants, these calculations are performed on reduced supercells, and the results are extrapolated to experimentally relevant defect concentrations using Matthiessen's rule (\autoref{sec:thermal}).

\begin{figure}[htbp]
\centering
\includegraphics[width=1.0\columnwidth]{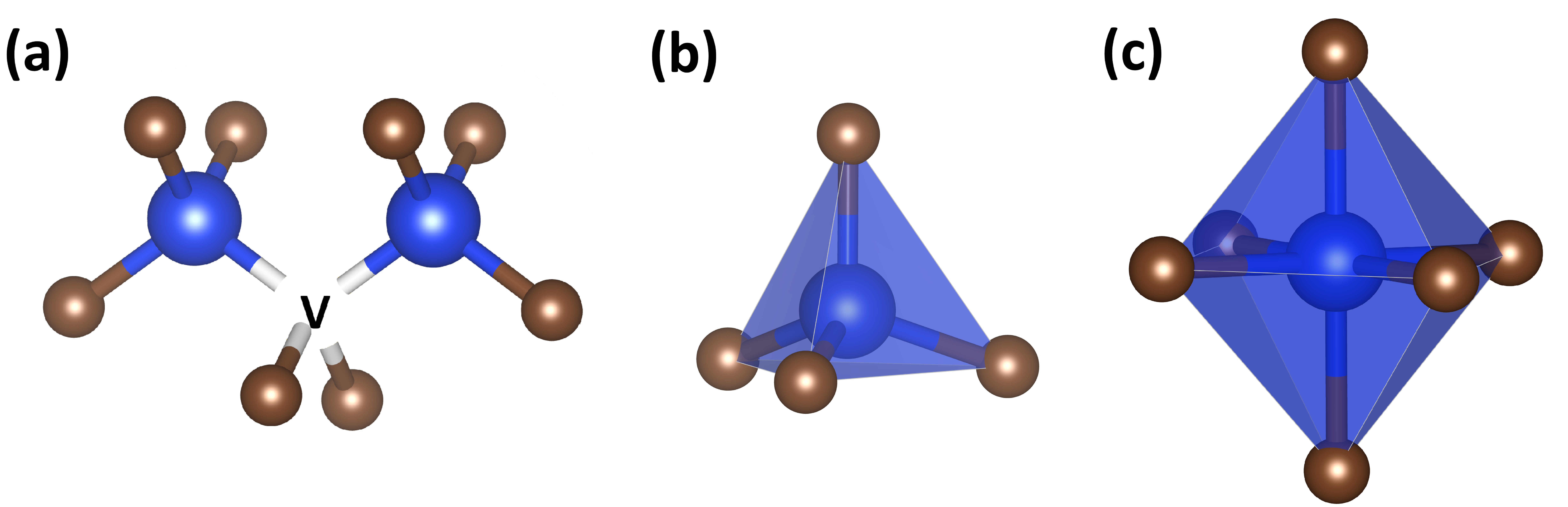}
\caption{A scheme representing local environments of considered defects: (a) boron-vacancy-boron (BVB) defect, (b) boron and phosphorus single substitutional defects, and (c) phosphorus-vacancy (PV) defect.
}
\label{fig:defects}
\end{figure}

\begin{figure}[htbp]
\centering
\includegraphics[width=1.0\columnwidth]{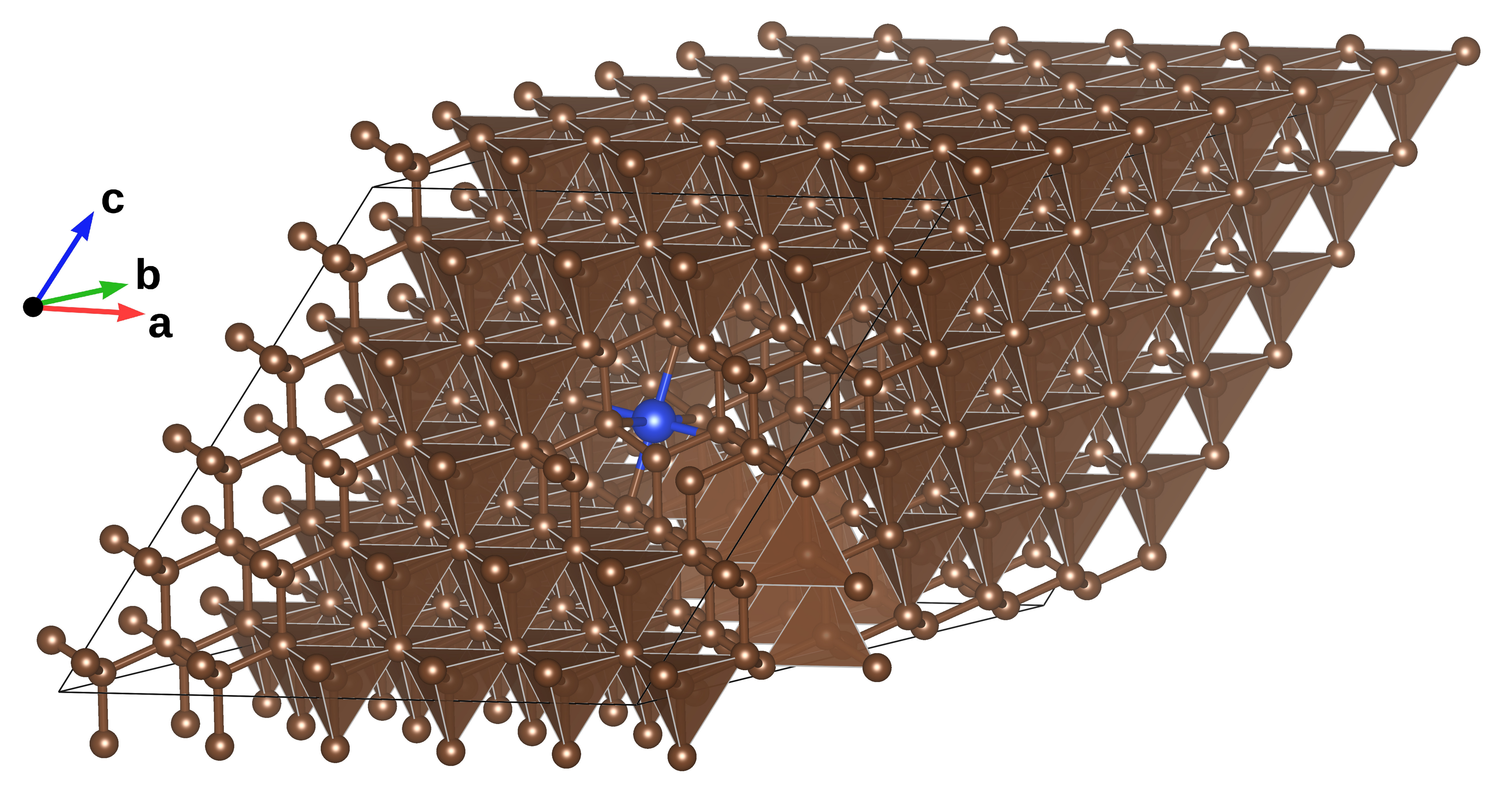}
\caption{An example of supercell doped with PV defect. The blue sphere represent the phosphorus atom, while the brown spheres indicate the positions of the carbon atoms.
}
\label{fig:supercell}
\end{figure}

As discussed in the introduction, we investigate two device architectures: a PIN and a PN junction.
The PIN architecture consists of boron-doped (P-region), boron-vacancy-boron-doped (BVB, I region), and phosphorus-doped (N-region) regions, while PN junction is formed by phosphorus-vacancy-doped (PV, P-region) and phosphorus-doped (N-region) regions (\autoref{fig:defects}).
To balance computational feasibility and realistic doping levels, a defect concentration of 0.4$\%$ is employed, corresponding to $5\times5\times5$ supercells of the primitive diamond unit cell.
This results in supercells containing 249 or 250 atoms, depending on the presence of a vacancy.
As an example, the 249 atom supercell doped with PV defect is shown in \autoref{fig:supercell}.

We performe all calculations without spin polarisation.
Our benchmark calculations indicate that spin polarisation is suppressed at the defect concentrations and temperatures relevant for device operation.
While the BVB defect is expected to exhibit paired spins \cite{PhysRevB.73.085204}, the situation is less straightforward for isolated substitutional dopants and PV complexes, where local magnetic moments may arise.
In particular, spin polarisation has been frequently associated with dopant-vacancy complexes \cite{PhysRevB.77.155206, PhysRevB.98.075208}; therefore, we explicitly benchmark its stability in the PV system.
To assess this behaviour, we analyse the dependence of the magnetisation on the electronic smearing, which controls the occupation of states near the Fermi level.
As shown in \autoref{tab:smearing_magnetisation}, increasing occupancy at finite temperature \cite{PhysRevB.65.035111} leads to a gradual suppression of the magnetic moment, reflecting the sensitivity of spin polarisation to partial occupations.
Furthermore, to explicitly evaluate the role of defect-defect interactions, we compare two systems: \textit{a)} a smaller supercell corresponding to 1.78\% defect concentration (53 atoms) with $\Gamma$-point sampling, which suppresses defect-state dispersion and thereby artificially enhances their localisation, and \textit{b)} a $2\times2\times2$ supercell of system \textit{a)} (424 atoms), explicitly containing multiple interacting PV defects, while preserving the same defect concentration.
The former exhibits a finite magnetic moment of 1.0 $\mu_B$, while the latter yields a non-magnetic solution.
This demonstrates that defect-defect interactions at high concentrations suppress spin polarisation.
Based on these results, we may conclude that the systems considered in this work are effectively non-magnetic at the relevant defect concentrations and temperatures above 180 K.

\begin{table}[htbp]
\centering
\setlength{\tabcolsep}{3pt}
\begin{tabular}{cc}
\textbf{Smearing Temperature} $T$ (K) & \textbf{Magnetisation} $M$ ($\mu_B$) \\
\hline
1   & 0.778 \\
10  & 0.778 \\
100 & 0.328 \\
120 & 0.308 \\
140 & 0.271 \\
160 & 0.205 \\
180 & 0.044 \\
200 & 0.000 \\
300 & 0.000 \\
\end{tabular}
\caption{Benchmark of total magnetisation as a function of Fermi-Dirac smearing temperature for the PV-doped system. The magnetic moment undergoes significant quenching as the electronic temperature exceeds 100 K.}
\label{tab:smearing_magnetisation}
\end{table}

To analyse transport through the depletion region, we further consider charged defects present in the PN junction, namely PV$^-$ and P$^+$.
The B$^-$ system relevant for the PIN junction is not explicitly considered, as the depletion region within the P and N layers is negligible compared to the width of the I absorber.
In semiconductor junctions, the depletion region associated with PV$^-$, P$^+$, and BVB systems is characterised by the presence of strong electric fields.
To assess the impact of such fields on the electronic structure and phonon dispersion, we perform benchmark calculations on representative systems consisting of substitutional boron in diamond at 6.25$\%$ concentration and pristine diamond.
We find that the application of an external electric field of approximately 10.7 MV/cm leads to a maximum change of 0.71 meV in the phonon eigenvalues, while the electronic eigenvalues are affected by less than 1.0 meV.
These variations are negligible on the energy scales relevant for transport and electron-phonon coupling.
Therefore, the explicit inclusion of the electric field in the charged-defect and BVB system calculations is not expected to significantly affect the results, and therefore, it is neglected.

\section{Results and discussion}
In this section, we analyse the suitability of the selected defect-engineered diamond structures for the proposed devices, both individually and as integrated systems such as PIN and PN junctions.
We begin with an analysis of defect formation energies, which provide qualitative insights into their relative thermodynamic stability and likelihood of formation.
This is followed by an examination of the defect geometries and electronic band structures, justifying their selection for the proposed applications.
In the detailed analysis of the electronic band structures, we give particular attention to the impact of the GW correction, especially on band gap sizes and the positions of impurity levels.
We then investigate the optical absorption of the relevant models, including the excitonic effects where applicable.
Subsequently, we analyse transport properties derived from electron-phonon coupling calculations, including carrier mobilities and electrical conductivities, and evaluate the lattice thermal conductivity.
Finally, we incorporate the first-principles results into an electrostatic (Poisson) solver to assess device-level behaviour and discuss the limitations of our study.

\subsection{Formation energies}
To provide preliminary guidelines for the experimental synthesis of diamonds containing the proposed defects, we calculate the formation energies for all considered systems.
Although more comprehensive insights could be obtained from phase diagram calculations at different temperatures and pressures, such simulations are beyond the scope of this study.
We point out that formation energies are often corrected for finite-size (supercell) effects; 
however, in this work we consider diamond-based materials containing defect concentrations corresponding exactly to those in the simulated supercells.
As a result, elastic and electrostatic finite-size corrections \cite{RevModPhys.86.253} are not applied, since we are not attempting to extrapolate the results to the dilute defect limit.

The formation energy \cite{RevModPhys.86.253, MURUGANATHAN2021108341, ZHANG2023109544,GAO2023109651} of a defect $D$ is computed as:
\begin{equation}
\label{eq:formation}
 E^f [D] = E[D] - E[\text{bulk}] - \sum_i n_i \mu_i 
\end{equation}
where $E[D]$ and $E[\text{bulk}]$ denote the total energy of the defective structure and the equivalent pristine diamond supercell, respectively,
the quantity $n_i$ represents the number of atoms of type $i$ removed ($n_i< 0$, carbon) or added ($n_i > 0$, dopants) to the structure, while $\mu_i$ denotes the corresponding chemical potentials. 
The chemical potentials are calculated from \textit{a)} pure diamond ($\mu_C$), \textit{b)} $\alpha$-rhombohedral boron ($\mu_B$), and \textit{c)} black phosphorus ($\mu_P$).  
We report the calculated formation energies in \autoref{tab:formation_energy}.
We observe that the single substitutional P dopant exhibits the highest formation energy of 6.7 eV, despite its common use as n-type dopant \cite{GROTJOHN2014129}.
In fact, the formation energy of the PV defect is lower than that of single substitutional P, which is consistent with trends reported for dopants with large atomic radii \cite{ PhysRevB.72.035214}.
In the case of the cluster defects (BVB and PV), it is valuable to evaluate the binding energy $E^b$ \cite{RevModPhys.86.253} of the cluster defect $D$, which indicates whether individual dopants or vacancies prefer to remain isolated ($E^b > 0$) or combine to form cluster defects ($E^b < 0$).
The binding energy is defined as follows:
\begin{equation}
\label{eq:binding}
 E^b [D] = E^f[D] - nE^f[X] - mE^f[V]
\end{equation}
where $n$ and $m$ denote the number of dopants (1 or 2) and vacancies (0 or 1) in the defect, respectively, while $E^f[X]$ and $E^f[V]$ are the formation energies of the corresponding single substitutional dopant and a vacancy in the supercells of equivalent size, respectively.

We find that binding energies of the BVB and PV defects are negative, indicating that the clustered configurations are energetically favoured.
Moreover, in our previous study on cluster defects in diamond \cite{PhysRevApplied.23.054029} we showed that BVB defect has a lower formation energy than the BV colour center, which is consistent with what reported in the literature.

Studies of defected semiconductors often calculate charge transition levels (CTLs) to describe the charge stability of point defects \cite{MURUGANATHAN2021108341, PhysRevB.98.075208}.
However, this formalism is strictly defined in the dilute limit, where individual defects are treated as isolated (i.e. non-interacting), giving rise to discrete, localised electronic levels \cite{RevModPhys.86.253}.
In the present work, we deliberately consider high defect concentrations, leading to defect-defect interactions.
As a result, electronic defect states acquire a finite dispersion and form impurity bands.
Depending on the system, these range from relatively narrow, still partially localised bands to impurity bands hybridised with host band edges, leading to degenerate semiconductor behaviour.
Under these conditions, the assumptions underlying the CTL formalism are no longer strictly satisfied. 
The defect states are not characterised by a single, well-defined energy, but by a distribution of energies across the Brillouin zone and their occupation becomes a collective property of the impurity band, rather than an isolated defect feature.
For these reasons, we do not employ the common CTL formalism. 
Instead, we directly analyse electronic band structures, which provide a comprehensive description of impurity band positions and widths, valence and conduction band edge modifications, and Fermi level positioning in the high-concentration regime.

\begin{table}[htbp]
\centering
\begin{tabular}{lcc}
\textbf{Defect} & \textbf{\shortstack{Formation Energy $E^f$ \\ (eV/def)}} & \textbf{\shortstack{Binding Energy $E^b$ \\ (eV/def)}} \\
\hline
BVB & 3.189 & -5.876 \\
B   & 1.096 & --- \\
P   & 6.409 & --- \\
PV  & 4.923 & -8.359 \\
\end{tabular}
\caption{Defect formation and binding energies (eV per defect) for the considered defects in diamond.}
\label{tab:formation_energy}
\end{table}

\subsection{Geometry and electronic structure including the GW correction}
\label{sec:geo_elst}
\subsubsection{Geometry}

The first step in our analysis is to examine how the defects modify the geometry of pristine diamond.
As expected, the introduction of a defect reduces the symmetry of the pristine structure.
The single dopant defects (B and P) retain the face-centred cubic Bravais lattice of pristine diamond but reduce symmetry to the $F\bar{4}3m$ space group ($\#$216).
In these cases, each dopant remains tetrahedrally coordinated with its four neighbouring carbon atoms, preserving the local $T_d$ point group symmetry.
In contrast, the PV defect forms a distorted octahedral coordination with its six neighbouring carbon atoms ($D_{3d}$).
In this octahedral coordination, the dopant-carbon distances are equal, while the bond angles deviate from the regular octahedral values.
This configuration is commonly referred to as a \textit{split-vacancy} configuration and frequently occurs in the case of impurities with large atomic radii \cite{Aharonovich2011,PhysRevB.88.235205}.
The resulting structure owns the symmetries of the trigonal space group $R\bar{3}m$ ($\#$166).
In the structure containing the BVB defect, each dopant coordinates three carbon atoms, while both dopants have identical local environments.
The coordination polyhedra of the dopants are nearly regular, as only one of the three dopant-carbon distances differs by $\sim10^{-2}$ \r{A} from the other two.
This results in a structure possessing the symmetry of the orthorhombic space group $Imm2$ ($\#$44).

The positively and negatively charged states preserve the space group of the corresponding neutral structures, while slightly modifying the dopant-carbon bond lengths.

\subsubsection{Electronic structure}
We continue by analysing the DFT and GW-corrected electronic band structures.
We focus particularly on the band gaps and the positions of impurity levels, comparing the two methods to quantify the impact of many-body corrections on the defect states.
The electronic band structures are computed along the standard piecewise linear path connecting the high-symmetry points in the irreducible Brillouin zone corresponding to the defect Bravais lattice \cite{SETYAWAN2010299}.
Since we consider systems that possess different lattices, we additionally calculate the band structures along the high-symmetry path of the face-centred cubic lattice as a common reference (see Supplementary Material Section I).

\begin{figure}[htbp]
\centering
\includegraphics[width=1.0\columnwidth]{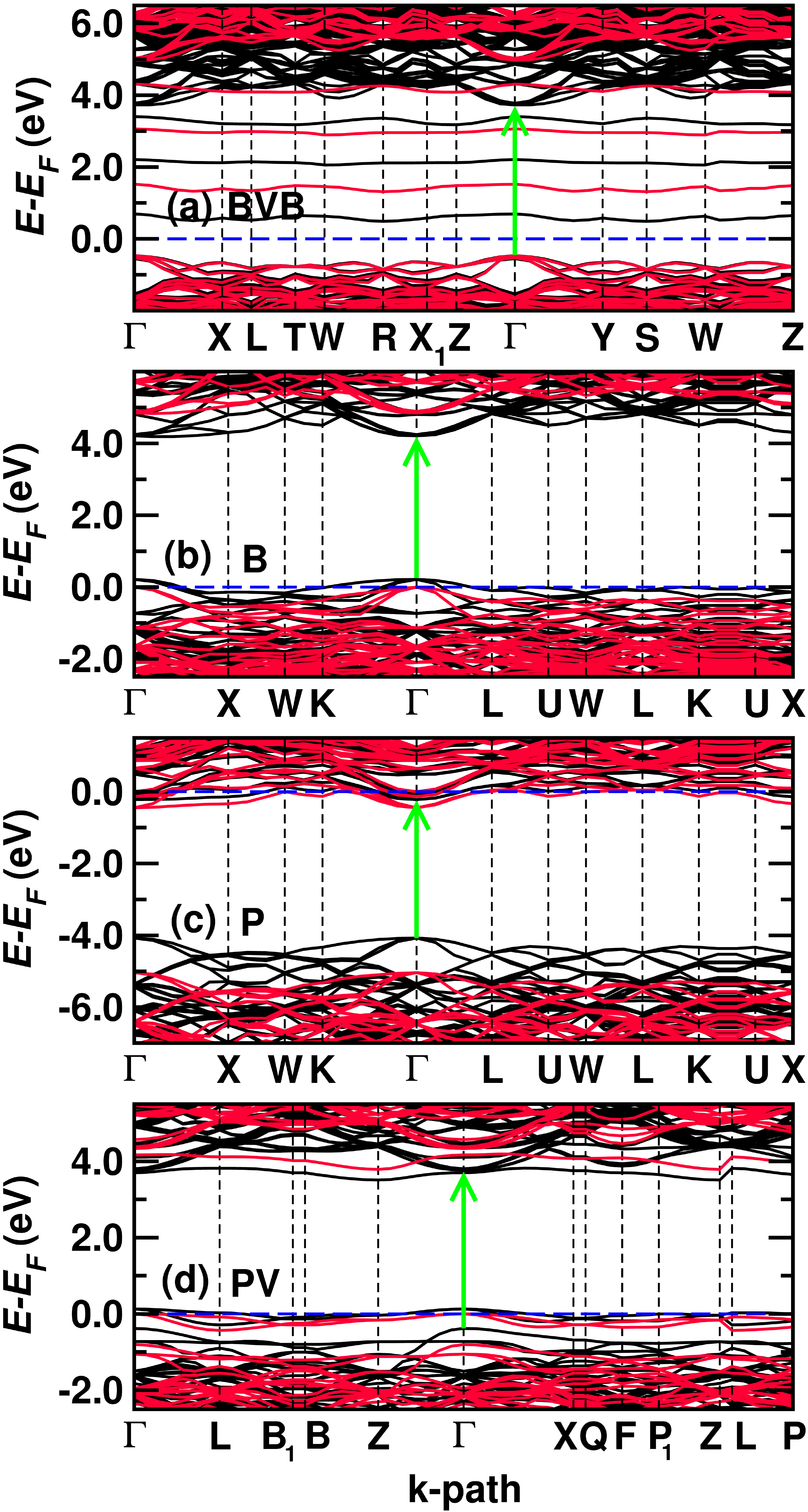}
\caption{Electronic band structures (BS) of system doped with (a) BVB, (b) B, (c) P, and (d) PV defects. The black and red lines correspond to the DFT and GW-corrected BS. The green arrows represent DFT band gaps reported in \autoref{tab:gw_corrections}. The Fermi level is set to 0 eV relative to the DFT BS and is indicated by a dashed blue line.
}
\label{fig:BS}
\end{figure}

The BVB defect introduces three impurity bands (IBs) within the diamond band gap with varying separations (\autoref{fig:BS}a), making it a promising candidate for intermediate-band photovoltaics.
Since all IBs are empty and located deep within the band gap, the structure behaves as an insulator.
The band gap of the BVB-doped diamond most closely resembles that of pristine diamond with a GW-corrected value of 5.44 eV, which is widened by 1.25 eV compared to the DFT value.
As expected, GW correction shifts all IBs upwards in energy and widens the band gap;
however, the crucial improvement arises from the changes in their relative positions.
In \autoref{tab:gw_corrections} we report relative energy differences (gaps) between the IBs.
It is clear that some energy separations increase significantly (VBM$\leftrightarrow$IB$_1$ and IB$_2$$\leftrightarrow$CBM), while others remain nearly unchanged (IB$_1$$\leftrightarrow$IB$_2$ and IB$_2$$\leftrightarrow$IB$_3$).

Our results show that impurity bands introduced by single substitutional boron and phosphorus defects merge with the valence and conduction band edges of diamond, respectively (\autoref{fig:BS}b,c).
This corresponds to a degenerate semiconductor regime; 
consequently, we expect metallic-like electronic behaviour, i.e. high electronic conductivity (see \autoref{sec:transport}).
The differences between the calculated band gaps and positions of selected impurities by the two methods are reported in \autoref{tab:gw_corrections}.
We observe that the degenerate semiconductor systems are the least affected by the GW correction, with the band gap increasing by 0.73-0.86 eV.
Due to the strong hybridisation between impurity and host states, distinguishing the exact onset of the host bands becomes non-trivial.
Therefore, for such systems we define the band gap as an energy difference between the valence band maximum (VBM) and the conduction band minimum (CBM), including the merged impurity bands rather than the intrinsic diamond host band gap (see green arrows in \autoref{fig:BS}b,c)).
The resulting band gap is effectively narrowed by two effects: \textit{a)} merging of the impurity levels with the host VBM or CBM, and \textit{b)} possible narrowing of the host band gap itself. 

The PV defect introduces two impurity bands within the host band gap: \textit{a)} an impurity band located above the VBM filled by three out of four electrons, and \textit{b)} an empty impurity band very close to the CBM (\autoref{fig:BS}d).
As it can be seen in \autoref{tab:gw_corrections}, the first impurity band is located $\sim$0.09 eV above the VBM according to the DFT result and $\sim$0.37 eV when GW corrections are applied.
This shift represents a significant difference when considering the activation energies of the defects.
The width of this impurity band is 0.42 (DFT)-0.43 (GW) eV and the Fermi level is located 0.3 eV above its bottom.
As a result, promoting an electron to the impurity band and creating a hole in the valence band requires approximately 0.67 eV, thus, at 300 K the number of free carriers in the valence band would be negligible. 
However, a significant width of this band suggests that it may support impurity-band metallic conduction, which we explore in \autoref{sec:transport} dedicated to the electronic transport. 
The GW correction also affects the second impurity band:
while DFT reports it as merged with the CBM, the GW locates it 0.14 eV below the CBM, while the host band gap is widened by $\sim$1 eV.

The considered structures doped with charged defects (P$^+$ and PV$^-$) represent insulating systems that can occur in the depletion regions of PN or PIN junctions.
Interestingly, we observe that DFT tends to slightly decrease the band gap sizes of the charged structures compared to their neutral counterparts, whereas GW produces the opposite effect by significantly widening the band gaps.
This results in GW shifts of 1.25-1.27 eV.
On the other hand, in the case of PV$^-$ the position of the first IB is 0.33 eV and 0.38 eV above the VBM according to DFT and GW, respectively, resulting in much closer agreement than in the case of the neutral PV defect.

We can conclude that GW calculation are essential for a more accurate determination not only of band gaps but also of the positions of impurity bands, which are crucial for estimating activation energies and optical absorption in materials suitable for electronic and  photovoltaic applications.
Moreover, the magnitude of the GW shifts appears to correlate with the electronic conductivity of the material: the largest shifts occur for insulating structures (BVB, P$^+$ and PV$^-$), the smallest for degenerate semiconductor structures (B and P), while the PV defect, which exhibits impurity-band conductivity lies in between.

\begin{table}[htbp]
\centering
\begin{tabular}{lccc}
\textbf{System} & \textbf{DFT (eV)} & \textbf{GW (eV)} & \textbf{$\Delta$ (eV)} \\
\hline
\multicolumn{4}{c}{\textbf{Band Gaps (CBM -- VBM)}} \\
\hline
B      & 3.98 & 4.83 & 0.86 \\
BVB    & 4.20 & 5.44 & 1.24 \\
P      & 3.86 & 4.59 & 0.73 \\
P$^+$  & 3.78 & 5.03 & 1.25 \\
PV     & 4.14 & 5.14 & 1.00 \\
PV$^-$ & 4.13 & 5.40 & 1.27 \\
\hline
\multicolumn{4}{c}{\textbf{PV Impurity Band Location}} \\
\hline
PV  & 0.09 & 0.37 & 0.28 \\
\hline
\multicolumn{4}{c}{\textbf{BVB Intermediate Energy Gaps}} \\
\hline
VBM$\leftrightarrow$IB$_1$ & 0.96 & 1.78 & 0.82 \\
IB$_1$$\leftrightarrow$IB$_2$ & 1.36 & 1.37 & 0.01 \\
IB$_2$$\leftrightarrow$IB$_3$ & 0.97 & 1.02 & 0.05 \\
IB$_3$$\leftrightarrow$CBM & 0.33 & 0.67 & 0.34 \\
\hline
\end{tabular}
\caption{Comparison of electronic energy gaps calculated using standard DFT and many-body perturbation theory (GW). 
The table presents the fundamental band gaps for various defect systems, the position of the first PV impurity band, and the energy spacings between the intermediate bands of the BVB defect. 
$\Delta$ represents difference between the DFT and GW reported values.}
\label{tab:gw_corrections}
\end{table}

\subsection{Absorption spectra}
The optical absorption of a photovoltaic device is one of its most critical characteristics.
It determines how effectively the material can harvest photons across the relevant spectral range, such as the solar spectrum at sea level (AM1.5G) or outside the Earth's atmosphere (AM0) \cite{GUEYMARD2002443}.
Moreover, it directly influences the required thickness of the absorbing material within the device, and thus device's cost, weight, and energy required for cell production.
In practice, photovoltaic cells (PN or PIN) often employ vertically stacked layers. 
Therefore, in the following, we analyse the optical properties of the absorbing material (the BVB-doped system) as well as those of potential top layers (boron or phosphorus-doped layers).
We further propose appropriate layer thicknesses and examine their effect on overall absorption.

\subsubsection{BVB absorber}

\begin{figure}[htb]
\centering
\includegraphics[width=1.0\columnwidth]{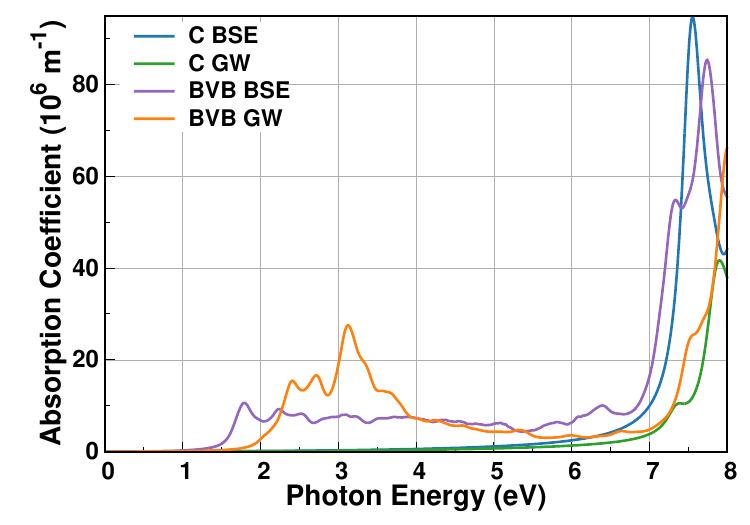}
\caption{Absorption coefficients along the \textit{x}-axis of the pristine (C) and BVB systems, calculated with BSE and GW-corrected RPA.
}
\label{fig:BSE}
\end{figure}

\begin{figure}[htb]
\centering
\includegraphics[width=1.0\columnwidth]{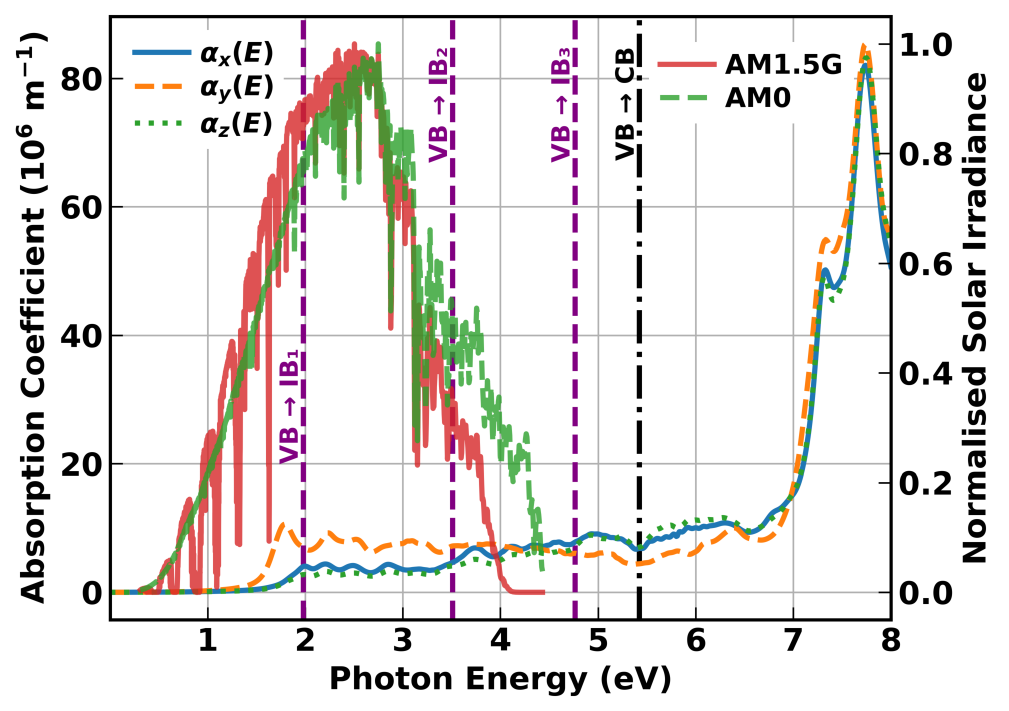}
\caption{Absorption spectrum $\alpha$ of the BVB absorber region along all three Cartesian directions compared to AM1.5G and AM0 solar spectra.
}
\label{fig:BVB_absorption_solar}
\end{figure}

We begin by analysing the excitonic effects on the absorption spectrum of the BVB absorber and compare them with those of pristine diamond.
To this end, we compare the GW-corrected Random Phase Approximation (RPA) and fully interacting Bethe-Salpeter equation (BSE) absorption spectra.
In the pristine system, we observe a redshift of the first sharp excitonic peak of approximately 0.33 eV in the BSE spectrum relative to the RPA result (\autoref{fig:BSE}).
This shift corresponds to the exciton binding energy and is consistent with previously reported values.
In contrast, the defected system exhibits an energy-dependent redshift.
The localisation of the intermediate-band (IB) defect states leads to significantly larger exciton binding energies \cite{RevModPhys.74.601}.
As a result, transitions involving these states are redshifted by approximately 0.6 eV.
Conversely, the excitonic effects for the perturbed bulk-like states are suppressed, resulting in a smaller redshift in the range of 0.25-0.30 eV.
Beyond these energetic redshifts, localisation alters the absorption peak intensities (oscillator strengths) \cite{RevModPhys.74.601}.
While the pristine system exhibits a stronger, sharper bulk transition peak in the BSE spectrum, the IB transitions in the BVB system are significantly weaker and broader compared to their relative RPA peak intensities. 
This suppression is also observed in the bulk-like states of the defected system, which show a significantly reduced ratio between the BSE and RPA peak intensities.
From a device perspective, the substantial 0.6 eV redshift of the IB transitions is highly advantageous. 
It shifts the sub-bandgap absorption deep into the 1.5 to 2.5 eV range, effectively aligning the defect's absorption profile with the peak photon flux of the AM1.5G and AM0 solar spectra, as seen in \autoref{fig:BVB_absorption_solar}.

\begin{figure}[htb]
\centering
\includegraphics[width=1.0\columnwidth]{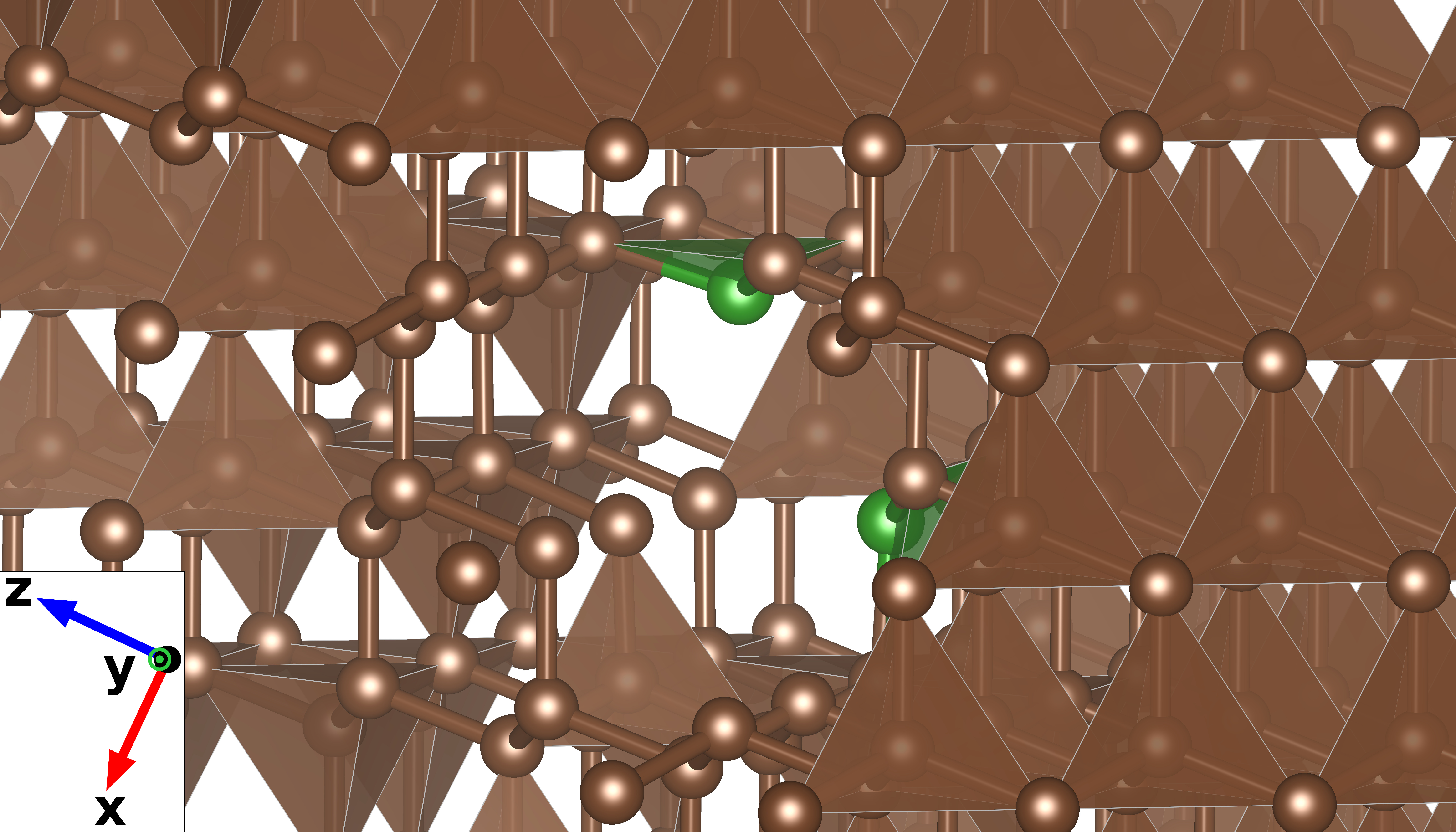}
\caption{Detail of the local environment of the BVB defect. Brown spheres and tetrahedra represent the bulk carbon lattice, while the green spheres mark the two boron atoms, with the vacancy positioned at the center of the local structure. The Cartesian axes define the directions used for anisotropic property analysis.
}
\label{fig:BVB_detail}
\end{figure}

\begin{figure}[htb]
\centering
\includegraphics[width=1.0\columnwidth]{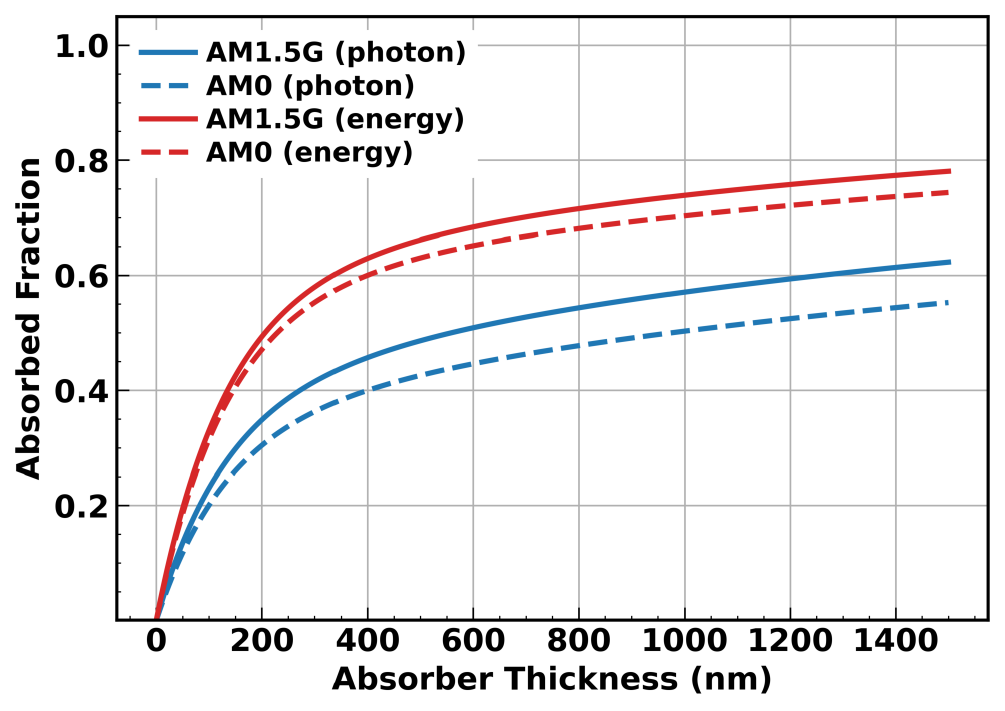}
\caption{Solar-weighted absorption in the BVB absorber as a function of layer thickness.
Solid lines correspond to AM1.5G illumination and dashed lines to AM0 conditions, while blue and red curves denote photon-weighted and energy-weighted absorption, respectively.
}
\label{fig:BVB_absorption}
\end{figure}

\begin{figure}[htb]
\centering
\includegraphics[width=1.0\columnwidth]{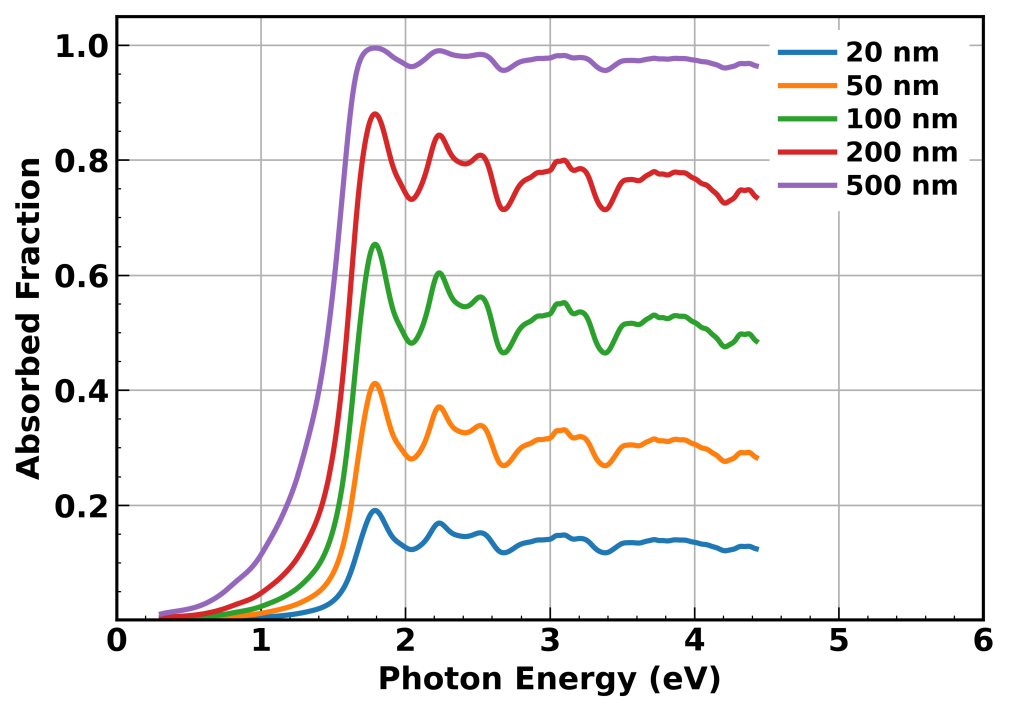}
\caption{Energy-resolved absorption in the BVB intermediate-band absorber for different layer thicknesses.
}
\label{fig:BVB_absorption_2}
\end{figure}

Now, we can proceed by analysing the ground state absorption properties of the absorber for photovoltaic applications.
As discussed above, the onset of defect-related absorption appears around 1.5 eV and forms a continuous, relatively constant absorption intensity up to the onset of bulk transitions at approximately 6.8 eV.
The orthorhombic symmetry of the defect (see defect orientation in \autoref{fig:BVB_detail}) induces anisotropy in the optical absorption of the defect states (\autoref{fig:BVB_absorption_solar}): absorption is maximised along the \textit{y}-axis in the energy range relevant to the solar spectrum, while \textit{x}- and \textit{z}-components become dominant above 4.1-4.6 eV range, i.e. above the VBM$\rightarrow$IB$_2$ and VBM$\rightarrow$IB$_3$ transition energies.
This behaviour can be attributed to the orbital character of the intermediate bands.
In particular, the IBs are primarily composed of states associated with boron dopants and their neighbouring carbon atoms, for IB$_1$ this corresponds to significant contributions from p$_x$, p$_y$, d$_{x^2-y^2}$, and d$_{xy}$ states, with a slight preference along the \textit{y}-direction.
In contrast, IB$_2$ and IB$_3$ exhibit predominantly z-oriented character, with contributions from p$_z$, d$_{z^2}$, d$_{yz}$, and d$_{xz}$ orbitals, leading to direction-dependent optical transitions.
Consequently, for standard photovoltaic operation, the optimal crystal orientation corresponds to incident light propagating within the \textit{xz}-plane of the simulated supercell, such that the electric field is aligned along the \textit{y}-direction, where the absorption is strongest.
In  \autoref{fig:BVB_absorption}, we show the total fraction of absorbed solar radiation as a function of absorber thickness, while \autoref{fig:BVB_absorption_2} presents the corresponding energy-resolved absorption spectra.
The absorption in the I region is described within the Beer-Lambert formalism:
\begin{equation}
A(E, d_I) = 1 - e^{-\alpha(E)\, d_I},
\end{equation}
where $A(E, d_I)$ is the fraction of photons with energy $E$ absorbed in an absorber of thickness $d_I$, and $\alpha(E)$ is the energy-dependent absorption coefficient of the I-region.
To evaluate the device-relevant absorption, this quantity is weighted over the solar spectrum.
The photon-weighted absorbed fraction is defined as
\begin{equation}
A_{\mathrm{ph}}(d_I) =
\frac{
\int \phi(E)\, \left[1 - e^{-\alpha(E)\, d_I}\right]\, dE
}{
\int \phi(E)\, dE
},
\end{equation}
where $\phi(E)$ is the incident solar photon flux as a function of energy $E$.
Alternatively, the energy-weighted absorption is given by
\begin{equation}
A_{\mathrm{en}}(d_I) =
\frac{
\int \phi(E)\,E\, \left[1 - e^{-\alpha(E)\, d_I}\right]\, dE
}{
\int \phi(E)\,E\, dE
},
\end{equation}
where the additional factor $E$ accounts for the energy carried by individual photons.
We find that increasing the absorber thickness beyond approximately 500 nm does not significantly enhance absorption, as the additional contribution arises mainly from energies below 1.5 eV, where the absorption coefficient is low.
This behaviour indicates that efficient absorption can be achieved with relatively thin absorber layers, which is advantageous for reducing material usage and fabrication costs.
It is also evident that the ground-state absorption onset occurs at higher energies than in conventional silicon solar cells (with an indirect band gap of 1.12 eV).
However, such a direct comparison neglects two key advantages of wide-bandgap intermediate-band solar cells: they can utilise high-energy photons more efficiently, and the present analysis considers only ground-state absorption, i.e. transitions from occupied states, while transitions from the intermediate bands to the conduction band are not included.



The natural next step beyond the ground-state analysis is to investigate the ladder-like transitions between intermediate bands (IB$_m$$\rightarrow$IB$_n$) and between IBs and the conduction band (IB$_m$$\rightarrow$CBM).
To this end, we calculate the optical absorption of excited states of the BVB absorber within the independent-particle approximation (IPA) without GW correction, as discussed in \autoref{sec:comp_details}.
For consistency, the ground-state absorption is also computed within the same framework as a reference.
We consider six different cases by promoting either 0.1 or one electron into IB$_1$, IB$_2$ and IB$_3$, respectively, followed by thermalisation of the hole and electron populations at 300 K.  
The 0.1-electron cases (denoted IB$_m^{(0.1)}$) represent systems with band structures very close to the ground state but with modified occupations, while the one-electron cases (denoted IB$_m^{(1.0)}$) correspond to a limiting regime under strong illumination, such as extreme concentrated photovoltaics, where the band structure is significantly altered.
In fact, in the one-electron cases, IB$_3$ merges with the conduction band, effectively reducing the number of distinct energy gaps and intermediate bands as summarised in \autoref{tab:excited_state_gaps}.

\begin{figure}[htbp]
\centering
\includegraphics[width=1.0\columnwidth]{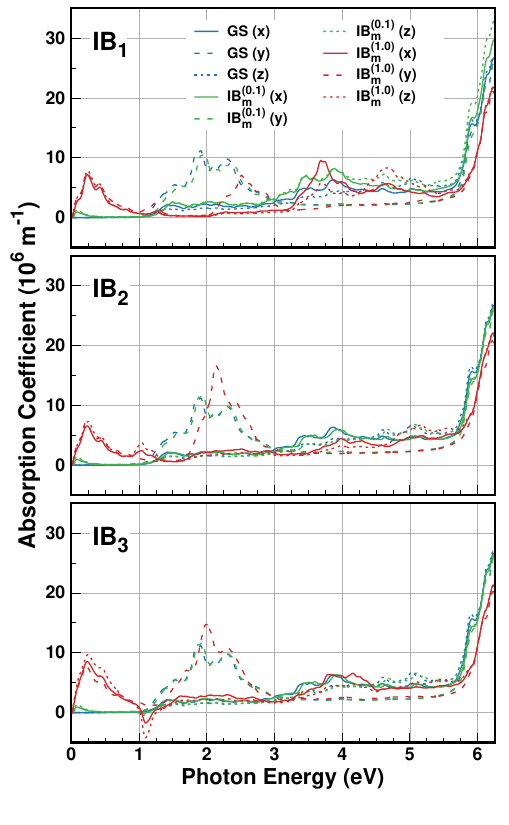}
\caption{Excited-state optical absorption spectra of the BVB absorber.
Panels show the directional response (\textit{x}, \textit{y}, and \textit{z} axes) upon promoting 0.1 or 1.0 electrons into IB$_1$ (top), IB$_2$ (middle), and IB$_3$ (bottom), compared against the ground state baseline.
Dashed lines: ground state  baseline, solid lines: IB$_m^{(0.1)}$, and dotted-line: IB$_m^{(1.0)}$.
Purple, green, and blue curves represent absorption along the \textit{x}-, \textit{y}-, and \textit{z}-directions, respectively.
}
\label{fig:excited_spectra}
\end{figure}

Upon partial occupation of IB$_1^{(0.1)}$, several optical effects emerge.
First, as in all considered excited-state systems, a low-energy peak appears near 0 eV (\autoref{fig:excited_spectra}).
This originates from intra-valence-band transitions as deeper electrons are optically excited into the newly created hole states at the VBM, leading to a slight blueshift of the overall absorption onset.
Moreover, populating IB$_1$ enhances the absorption along x- and y-directions not only at energies corresponding to IB$_1$$\rightarrow$IB$_m$ and IB$_1$$\rightarrow$CB transitions, but across the entire spectrum.
Interestingly, occupation of IB$_1$ also leads to near-degeneracy of the absorption spectra along the \textit{x}- and \textit{y}-directions for energies below 4 eV.
This behavior suggests a potential improvement in the absorber performance under steady-state illumination compared to predictions based solely on ground-state (GS) absorption.
Promotion of one electron per supercell leads to a substantial modification of the defect-related absorption compared to the ground state.
In addition to the intra-valence-band absorption peak, the spectrum is characterised by three prominent peaks corresponding to VB$\rightarrow$ IB$_m$ transitions, and a less pronounced peak associated with the IB$_1$$\rightarrow$IB$_2$ transition at approximately 1.26 eV.
As a result, the average absorption coefficient is lower, and less spread, mainly focused in such peaks.
Furthermore, since the band the intermediate bands are shifted to the higher energies, the overlap with the solar spectra is much lower. 
Interestingly, these prominent peaks exhibit their maxima along different Cartesian directions, highlighting the anisotropy arising from the spatial character of the intermediate-band states.

IB$_2^{(0.1)}$ case yields optical spectra very similar to the ground state, with the exception of a slight blueshift and a low-energy peak common to all consider excited-state systems.
A modest enhancement of absorption is observed around 1.9 eV, corresponding to the IB$_2$$\rightarrow$CBM transition, while minor improvement appears around 1.1 eV associated with the IB$_2$$\rightarrow$IB$_3$ transition.
A single peak located approximately at 2.1 eV dominates the optical absorption of a IB$_2^{(1.0)}$ system, corresponding to the VB$\rightarrow$IB$_1$ transition, while lesser prominent peak can be observed at $\sim$1 eV (IB$_2$$\rightarrow$CB).
In this system, the IB$_3$ has the highest overlap with the CB among all systems, while VB$\rightarrow$IB$_1$ transition requires less energy compared to IB$_1^{(1.0)}$ as reported in \autoref{tab:excited_state_gaps}.

The IB$_3$$\rightarrow$CB energy gap is the smallest in the BVB absorber and is the only one capable of utilizing photons with energies significantly below the band gap of silicon (see GW corrected gaps in \autoref{tab:gw_corrections})
Consequently, the analysis of the IB$_3^{(0.1)}$ system is of particular importance.
Indeed, absorption in the 0.4-0.6 eV range is increased compared to the ground state;
however, it remains negligible compared to the VB$\rightarrow$IB$_m$ transitions with an onset around 1.1 eV.
This is due to the significantly lower number of available electrons in the partially occupied intermediate band compared to the valence band.
Similarly to IB$_2^{(0.1)}$ case, we observe enhanced absorption in the 1.85-2.2 eV range when compared to ground state, which is particularly relevant as it coincides with the maximum intensity of the solar spectrum.

In both IB$_3^{(0.1)}$ and IB$_3^{(1.0)}$, negative absorption coefficients are observed, corresponding to stimulated emission due to population inversion.
Notably, even at high IB$_3$ occupation, this effect appears only along the \textit{x}- and \textit{z}-directions.
Moreover, the IB$_3^{(1.0)}$ absorption spectrum begins to resemble both the ground-state and IB$_3^{(0.1)}$ spectra, due to the reduction in the VB $\rightarrow$ IB$_1$ transition energy.   

In summary, the analysis of excited-state absorption indicates an overall improvement of optical absorption under steady-state illumination. 
Lower occupations of the intermediate bands, which do not significantly alter the band structure, generally yield more favorable absorption characteristics, particularly in the low-energy range relevant to the solar spectrum. 
In contrast, higher occupations shift the intermediate bands toward the conduction band, reducing overlap with the solar spectrum. 
This blueshift is particularly pronounced in the IB$_1^{(1.0)}$ case, while IB$_2^{(1.0)}$ and IB$_3^{(1.0)}$ exhibit spectra that remain more effective for solar energy harvesting.
Lower occupations are also expected to be more representative of standard photovoltaic operating conditions.
Furthermore, partial occupation of IB$_1$ (0.1 electrons) has the strongest impact on enhancing absorption, while at higher occupations the directional dependence of absorption becomes more pronounced, with distinct peaks emerging along different directions.
While partial occupation of the intermediate bands enhances absorption at higher energies, it does not significantly improve absorption in the low-energy region where the ground-state absorption is nearly zero (below 1 eV in \autoref{fig:excited_spectra}), except for intra-valence-band transitions, which are not useful for photovoltaic applications.   
To effectively utilise this low-energy region, either thicker absorber layers would be required compared to those proposed based on ground-state analysis, or significantly higher intermediate-band occupation would be needed, which may lead to undesirable shifts in the band structure, particularly in the case of IB$_1$, leading to decrease in absorption in lower energy regions..

\begin{table}[htbp]
\centering
\setlength{\tabcolsep}{3pt} 
\small
\begin{tabular}{lccccccc}
\textbf{Gap} & \textbf{GS} & \textbf{IB$_1^{(0.1)}$} & \textbf{IB$_2^{(0.1)}$} & \textbf{IB$_3^{(0.1)}$} & \textbf{IB$_1^{(1.0)}$} & \textbf{IB$_2^{(1.0)}$} & \textbf{IB$_3^{(1.0)}$} \\
\hline
VB$\rightarrow$IB$_1$     & 1.074 & 1.150 & 1.121 & 1.105 & 1.944 & 1.529 & 1.278 \\
IB$_1$$\rightarrow$IB$_2$ & 1.382 & 1.354 & 1.408 & 1.422 & 1.032 & 1.643 & 1.654 \\
IB$_2$$\rightarrow$IB$_3$ & 1.046 & 1.030 & 1.043 & 1.054 & 0.847 & 0.796 & 1.042 \\
IB$_3$$\rightarrow$CB     & 0.328 & 0.296 & 0.254 & 0.246 & ---   & ---   & ---   \\
\end{tabular}
\caption{Evolution of intermediate optical gaps (eV) with the change of occupation ($f$) of specific defect band IB$_m$. Values are shown for the ground state (GS) and partial occupations 0.1 and 1.0 of IB$_1$, IB$_2$, and IB$_3$.}
\label{tab:excited_state_gaps}
\end{table}

\subsubsection{Degenerate P- and N-region (B- and P-doped)}

We proceed by analysing the absorption spectra of the P- and N-region (boron- and phosphorus-doped) of the photovoltaic device.
In this section, we focus on how these regions may attenuate the solar radiation reaching the absorber layer, as they often form the top layer of the device.
In general, their absorption should be minimized, since their low-energy absorption (below the onset of bulk-state transitions) arises from intra-valence-band processes in the P-region and intra-conduction-band processes in the N-region, and is therefore not useful for photovoltaic applications.
We recall that boron- and phosphorus-doped structures exhibit face-centred cubic symmetry, resulting in isotropic absorption along the \textit{x}-, \textit{y}-, and \textit{z}- directions.

\begin{figure}[htb]
\centering
\includegraphics[width=1.0\columnwidth]{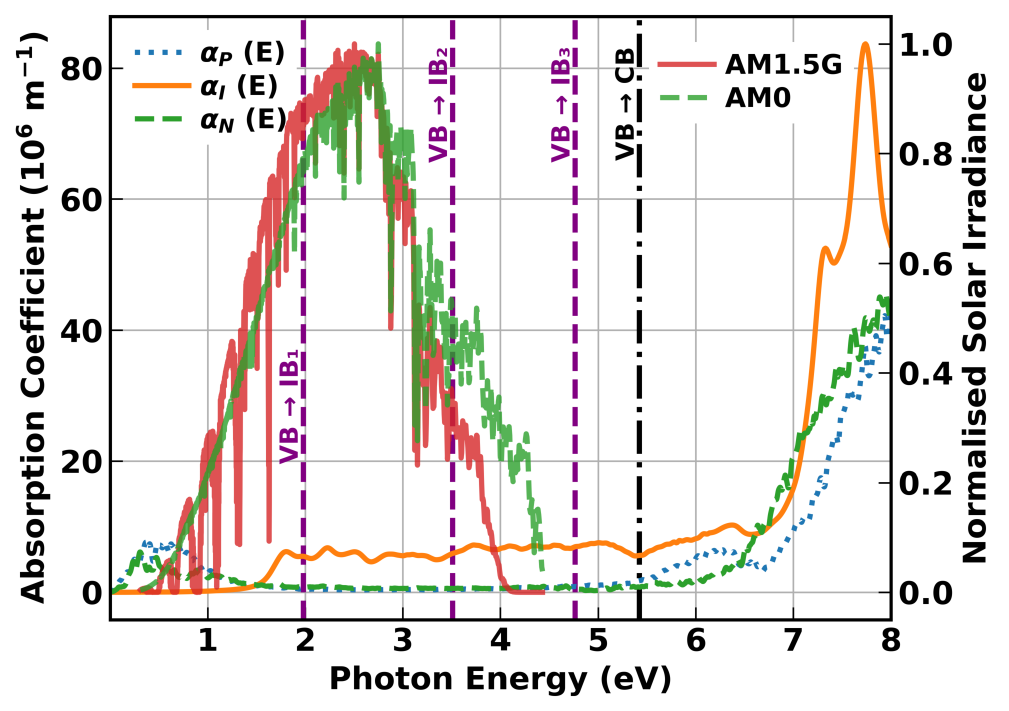}
\caption{Absorption spectrum $\alpha$ of the P and N layers compared to BVB absorber layer (directional average) and AM1.5G and AM0 solar spectra.
}
\label{fig:PN_spectra}
\end{figure}

\begin{figure}[htb]
\centering
\includegraphics[width=1.0\columnwidth]{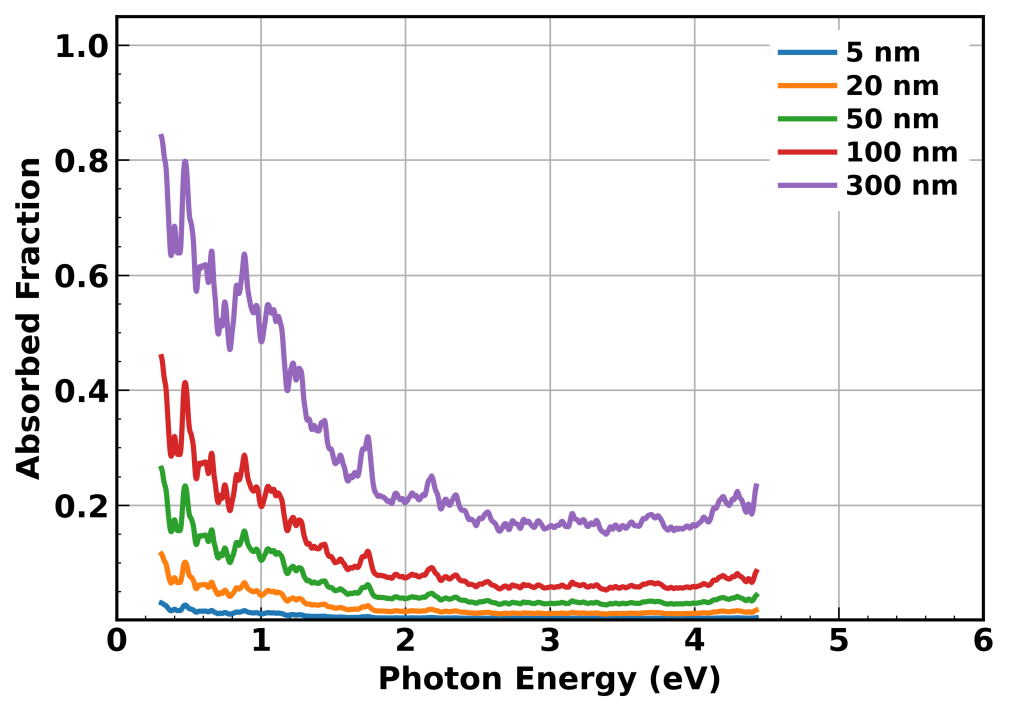}
\caption{Energy-resolved loss in the N-layer (phosphorus-doped) of different thicknesses.
}
\label{fig:N_spectra}
\end{figure}

\begin{figure}[htb]
\centering
\includegraphics[width=1.0\columnwidth]{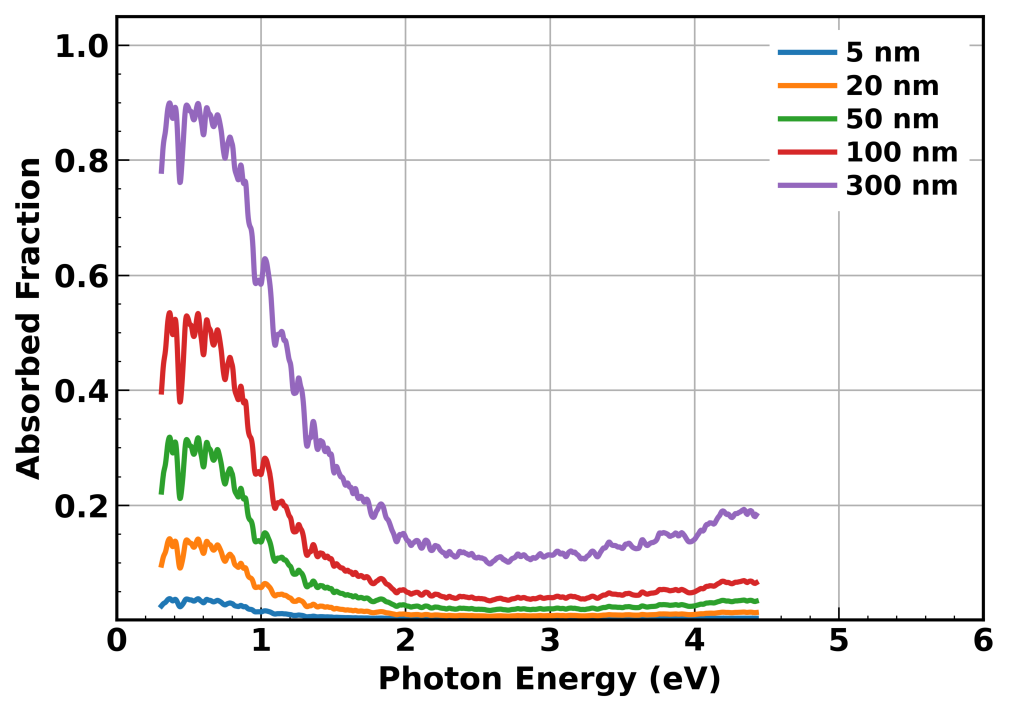}
\caption{Energy-resolved loss in the P-layer (boron-doped) of different thicknesses.
}
\label{fig:P_spectra}
\end{figure}

As shown in \autoref{fig:PN_spectra}, the absorption of both P- and N-layer lies mostly outside the main solar spectral range.
Furthermore, the low-energy intra-band absorption occurs outside the energy range where the BVB absorber is active, and therefore does not directly compete with it.
We observe that the P-layer exhibits stronger long-wavelength intra-band absorption; however, in the relevant 1.5-4.4 eV energy window, its absorption is up to 30$\%$ lower than that of the N-layer.
Consequently, from an optical perspective, the P-layer appears to be a more suitable choice for the top layer of a monofacial PIN photovoltaic device.
The fractions of photons absorbed in the N- and P-layers are reported in \autoref{fig:N_spectra} and \autoref{fig:P_spectra}, respectively.
In the relevant energy range, even a thickness of 100 nm for the P- and N-layers does not significantly reduce the overall absorption, suggesting that bifacial device configurations may be feasible.


\subsection{Electron-phonon coupling - transport}
\label{sec:transport}
We recall that our goal is to analyse two distinct device architectures: a PIN junction with an intermediate-band absorber, and a PN junction formed purely by phosphorus doping, where the P-region is created via phosphorus-vacancy (PV) complexes.
A crucial characteristic of any semiconductor junction device is its electronic transport.
The carrier mobility of the absorber determines how efficiently carriers move through the material and whether they can be extracted before recombination.
In addition, the conductivity of the degenerately doped B- and P-doped regions, as well as the impurity-band-conducting PV-doped system, governs their suitability as contact layers, influencing the series resistance of the device and the formation of ohmic contacts with metals.
To quantify these effects, we calculate electron-phonon-coupling-related transport properties, including phonon-limited carrier mobilities, electrical conductivities, electronic thermal conductivity, and Seebeck coefficients.
All transport properties in the following discussion are evaluated within the momentum relaxation time approximation (MRTA) \cite{Ponce_2020}.
It should be noted that, since the electron-phonon coupling is derived directly from the defected structures, the present formalism naturally captures the complex scattering arising from local perturbations introduced by the dopants.
However, additional Coulomb scattering \cite{10.5555/1594006} from ionised impurities is not explicitly included.
As a result, the calculated mobilities represent upper bounds on carrier transport.

\subsubsection{BVB absorber}
\begin{figure}[htb]
\centering
\includegraphics[width=1.0\columnwidth]{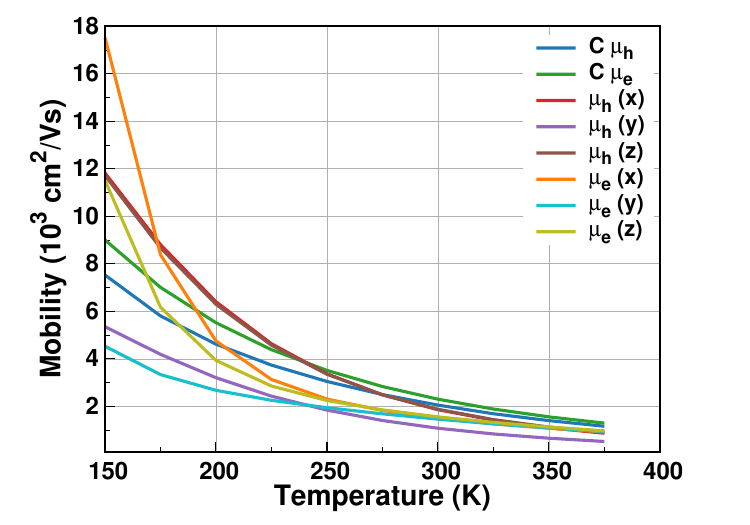}
\caption{Hole and electron mobility ($\mu$) as a function of temperature of the BVB absorber and pure diamond (C).
}
\label{fig:mobility_BVB}
\end{figure}

We begin by analysing the BVB-doped system, which serves as the light-absorbing layer in the device.
We recall that the electron-phonon coupling was calculated for a smaller BVB-doped supercell, as discussed in \autoref{sec:comp_details}.
In this system, we are primarily interested in how the presence of the defect affects the mobility of holes and electrons near the valence band maximum (VBM) and conduction band minimum (CBM), respectively, compared to pristine diamond.
As shown in \autoref{fig:mobility_BVB}, the introduction of the BVB complex significantly alters the transport properties. 
First, the reduction of symmetry due to the defect induces strong anisotropy, leading to a splitting of the macroscopic mobilities along the principal Cartesian directions. 
Interestingly, at low temperatures (e.g. 150 K) the defected structure exhibits significantly enhanced mobility along specific directions compared to pristine diamond.
Most notably, the electron mobility exhibits a large directional spread, reaching a maximum of $\sim$17,500 cm$^2$ V$\cdot$s along the \textit{x}-direction, while simultaneously showing the lowest mobility in the system along the \textit{y}-direction ($\sim$4,500 cm$^2$/(V$\cdot$s)).
In addition, the hole mobility along the \textit{x}- and \textit{z}-directions is also enhanced.
This suggests that the defect modifies both the valence and conduction bands curvature, significantly reducing the effective mass along these directions.
However, as the temperature increases, the phonon-limited mobility of the BVB system decreases much more rapidly than that of pristine diamond.
Moreover, the rate of this decrease depends on the transport direction, indicating strongly anisotropic electron-phonon coupling.
The highly mobile electrons along the \textit{x}-direction exhibit the steepest temperature dependence, suggesting stronger electron-phonon coupling along this axis.
In contrast, the electron mobility along the \textit{y}-direction decreases more gradually.
As a result, above approximately 325 K, the mobility along the \textit{y}-direction exceeds that along the \textit{x}-direction, and the electron mobilities become nearly isotropic around 375 K.
In contrast, the hole mobilities retain a significant anisotropic spread across the entire temperature range.
At 300 K, the hole mobility remains comparable to that of pristine diamond along the \textit{x}- and \textit{z}-directions ($\sim$1900 cm$^2$/Vs), while it is reduced to $\sim$1080 cm$^2$/Vs along the \textit{y}-direction.
Similarly, the room-temperature electron mobilities converges to a nearly isotropic $\sim$1400-1500 cm$^2$/(V$\cdot$s).
Although the introduction of the BVB complex reduces the overall room-temperature mobility relative to pristine diamond, the resulting values remain high, highly exceeding those of common thin-film photovoltaic absorbers such as CdTe \cite{10.1063/1.4984320} or metal-halide perovskites \cite{Stranks2015}.
Since these properties were calculated for a smaller supercell (127 atoms), the corresponding values for the larger (249-atom) supercell are expected to be slightly closer to those of pristine diamond, while preserving the same qualitative trends.

\subsubsection{Degenerate P- and N-region (B- and P-doped)}

In addition to enabling efficient carrier transport, the degenerate P- and N-type regions serve a dual function in the device by establishing the built-in electric field of the junction while simultaneously providing highly conductive pathways for carrier extraction.
We recall that in degenerate semiconductors the Fermi level lies within the valence band (P-type) or conduction band (N-type), and the transport properties are therefore governed by electronic states at the Fermi level rather than by those near the valence and conduction band edges.
While the built-in electric field governs carrier separation, the electrical conductivity determines how efficiently carriers are transported and extracted. 
This is particularly relevant when comparing PIN and PN architectures: in PIN devices, the field extends across the intermediate-band absorber, and the degenerate regions primarily act as conductive contacts, whereas in PN devices, transport through the doped regions plays a critical role, making high conductivity essential for efficient device operation.

\begin{figure}[htb]
\centering
\includegraphics[width=1.0\columnwidth]{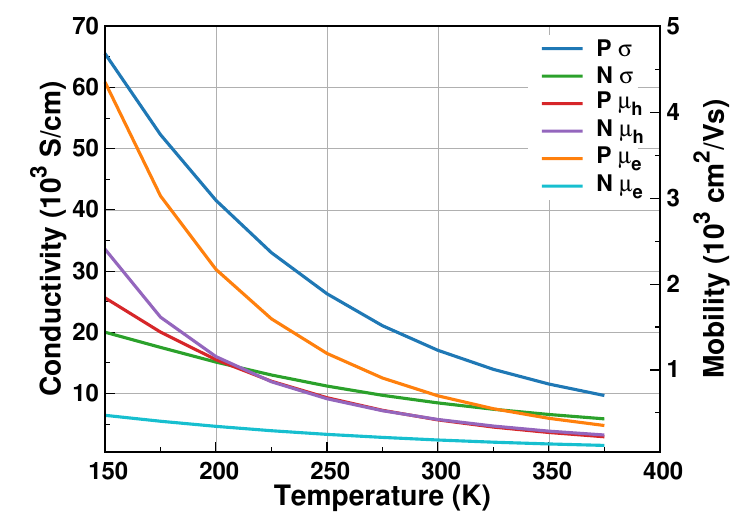}
\caption{Electronic conductivity ($\sigma$) and electron/hole mobility ($\mu$) in the degenerately doped P- and N-region as a function of temperature.
}
\label{fig:deg_conductivity}
\end{figure}

As shown in \autoref{fig:deg_conductivity}, the P and N type regions reach phonon-limited electrical conductivities of $\sim$17,000 S/cm and $\sim$8,500 S/cm at 300 K, respectively, making them highly conductive.
These high conductivities arise from the large carrier concentrations combined with relatively high carrier mobilities.
At 300 K, the P-region exhibits hole and electron mobilities of $\sim$420 and $\sim$700 cm$^2$/Vs, respectively, while the N-region shows corresponding values of $\sim$423 and $\sim$185 cm$^2$/Vs.
The position of the Fermi level within the bands, the high conductivity values, and the decrease of conductivity with increasing temperature consistently indicate a degenerate, metal-like transport regime in both systems.
As a result, the proposed boron- and phosphorus-doped materials are viable candidates for the P- and N-type regions of the device, as they provide sufficiently low series resistance, support the formation of ohmic-like contacts at the metal-semiconductor interface, and enable the establishment of a strong built-in electric field across the junction.

\subsubsection{PV-doped P Region}

As we discussed in the introduction, one of our objectives is to assess the feasibility of a PN junction based on a single dopant atomic species. 
Since phosphorus-vacancy doped diamond exhibits p-type behaviour, this provides an opportunity to combine it with substitutional phosphorus in a PN junction.
As further discussed in \autoref{sec:geo_elst}, the Fermi level within the impurity band is located approximately 0.67 eV above the valence band maximum, making thermal excitations of electrons from the valence band into the impurity band nearly negligible at room temperature.
Consequently, charge transport is governed by carriers within the impurity band itself.
This transport regime is naturally captured in our electron-phonon coupling calculations, as the Fermi level crosses a relatively broad impurity band.

\begin{figure}[htb]
\centering
\includegraphics[width=1.0\columnwidth]{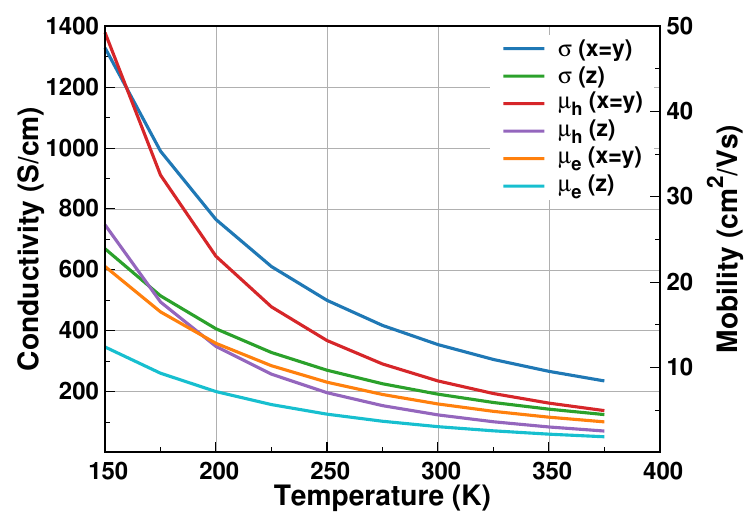}
\caption{Electronic conductivity ($\sigma$) and electron/hole mobility ($\mu$) in the PV-doped system as a function of temperature.
}
\label{fig:PV_conductivity}
\end{figure}

We observe that the transport properties are identical along \textit{x}- and \textit{y}-directions, while those along the \textit{z}-direction are approximately half as large (\autoref{fig:PV_conductivity}).
At 300 K, the electrical conductivity reaches approximately 355 S/cm along the \textit{x}- and \textit{y}-directions and 190 S/cm along the \textit{z}-direction.
This behaviour, together with the decrease of conductivity with increasing temperature, indicates a degenerate transport regime similar to that observed in the boron- and phosphorus-doped systems.
However, the overall conductivity is approximately 50-100 times lower than in the boron-doped case.
This reduced mobility reflects the flatter dispersion and increased scattering in impurity-band states compared to states in the valence band.
At 300 K, the hole and electron mobilities reach approximately 8.4 and 5.7 cm$^2$/Vs along the \textit{x}- and \textit{y}-directions, respectively, while along the \textit{z}-direction they are further reduced to approximately 4.5 and 3.1 cm$^2$/Vs
From a device perspective, this level of conductivity remains sufficiently high for practical operation, making the material a promising degenerate p-type semiconductor for applications in tunnel diodes or PIN junctions.
On the other hand, it is important note that if the impurity-band contribution is suppressed (i.e. by reducing the defect concentration), the material becomes effectively insulating.
Consequently, the use of PV-doped material for conventional rectifying PN junction is limited.

\begin{figure}[htb]
\centering
\includegraphics[width=1.0\columnwidth]{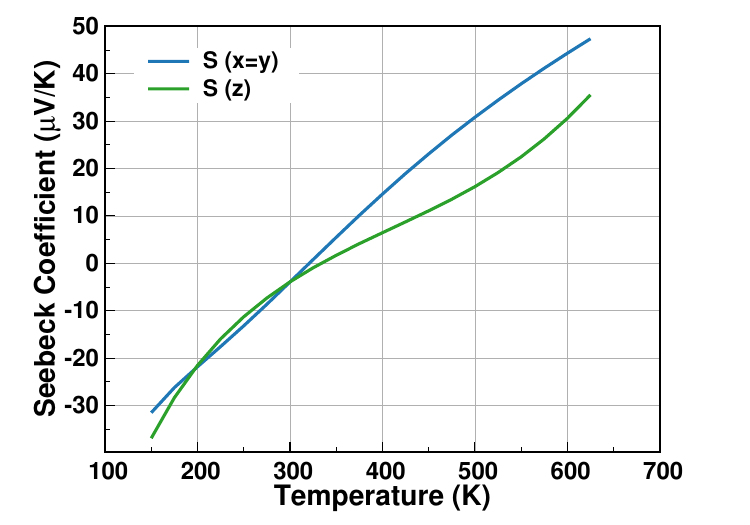}
\caption{Sebeck coefficient (S) of the PV-doped system as a function of temperature.
}
\label{fig:PV_seebeck}
\end{figure}

Furthermore, calculations of the Seebeck coefficient reveal anisotropic and bipolar transport behaviour within the PV-induced impurity band (\autoref{fig:PV_seebeck}).
At low temperatures, the material exhibits a negative Seebeck coefficient across all axes, indicative of electron-dominated transport. 
However, as temperature increases, thermal excitation drive a transition to positive, hole-dominated transport. 
Notably, this bipolar crossover occurs at different temperatures along different Cartesian directions. 
Around 325 K, the material exhibits an unusual goniopolar transport regime, in which it behaves as a p-type conductor in the $xy$-plane ($S_x > 0$) while simultaneously exhibiting n-type behaviour along the $z$-axis ($S_z < 0$).
Such anisotropy and sign-reversal of the Seebeck coefficient may be of interest for applications where thermoelectric voltages need to be minimised or controlled, for example in interconnects operating across temperature gradients.
Because the Seebeck coefficient spans both negative and positive values as a function of temperature, the resulting thermovoltage ($V = \int S(T)\; dT$) \cite{kittel} can exhibit partial compensation over a given temperature range.
By selecting appropriate crystal orientations, it may therefore be possible to engineer vanishing or strongly reduced thermovoltages for specific temperature pairs.
For example, a near-zero net thermovoltage can be achieved for a temperature difference between $T_\mathrm{cold} \sim$150 K and $T_\mathrm{hot} \sim$493 K along the $xy$-plane.
Beyond interconnects, the sign-changing thermopower opens pathways toward active thermal control devices designed to keep a specific element exactly at the $S=0$ crossing temperature.
Because the Peltier coefficient $\Pi = S \cdot T$ changes sign at the Seebeck crossover temperature, the direction of Peltier heating and cooling reverses across this point.
Under applied current, this can give rise to a self-compensating thermoelectric response, where deviations from the crossover temperature induce opposing heating or cooling contributions.
This behaviour suggests the possibility of passive thermal stabilisation around the $S=0$ crossover temperature.
Finally, the position of the Seebeck sign crossover, as well as the magnitude of the thermopower, can in principle be tuned through defect concentration or by combining different crystallographic orientations within a single device.

\subsubsection{PV$^-$ and P$^+$ charged defects}

To approximate carrier transport in the depletion region of a PN junction formed by phosphorus-vacancy- and phosphorus-doped regions, we calculate the transport properties of their charged counterparts (PV$^-$ and P$^+$).
As mentioned in \autoref{sec:comp_details}, the B$^-$ system is not considered, since in the PIN junction the depletion region within the P and N layers is negligible compared to the width of the I absorber.
The charged systems are insulating; therefore, the carrier mobilities are evaluated at the valence, conduction, or impurity band edges, in contrast to the neutral degenerate systems, where transport is governed by states at the Fermi level.
Moreover, the additional charge modifies the local environment of the defect and induces slight changes in the band structure.

\begin{figure}[htb]
\centering
\includegraphics[width=1.0\columnwidth]{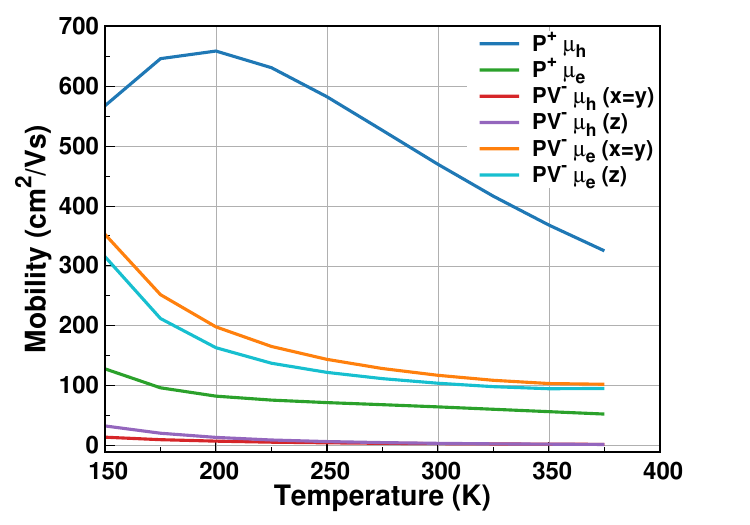}
\caption{Electron/hole mobility ($\mu$) in the charged systems as a function of temperature.
}
\label{fig:charged_mobility}
\end{figure}

We observe that the mobilities of both carrier types are reduced in the P$^+$ system compared to the neutral P-doped structure (\autoref{fig:charged_mobility}).
At 300 K, the hole mobility reaches approximately 470 cm$^2$/Vs, significantly exceeding the electron mobility of approximately 65 cm$^2$/Vs.
This can be attributed to the fact that the conduction band is more strongly affected by the P$^+$ defect than the valence band.
We recall that the PV defect introduces two impurity bands into the band structure: one located above the valence band, responsible for impurity-band transport in the neutral case, and a second one positioned close to the conduction band.
Consequently, in the charged PV$^-$ system, hole mobility is evaluated at the top of the lower impurity band, while electron mobility is determined at the bottom of the higher-energy impurity band.
Interestingly, relatively high electron mobility is observed, reaching approximately 117 cm$^2$/Vs in the $xy$-plane and 104 cm$^2$/Vs along the $z$-direction at 300 K.
In contrast, the hole mobility is very low, reaching approximately 3.2 cm$^2$/Vs and 3.7 cm$^2$/Vs in the $xy$-plane and along the $z$-direction, respectively.
Notably, the highest mobility is found along the \textit{z}-direction, in clear contrast with the neutral case.
Overall, these results suggest that depletion region does not significantly worsen the transport compared to neutral regions.
In particular, for complex defects such as PV, the available conduction channels associated with impurity bands may even support more efficient carrier transport.
Moreover, under operating conditions, the presence of a strong built-in electric field ensures that carriers traverse the depletion region rapidly, reducing the impact of low-mobility regimes on overall device performance.

\subsection{Lattice thermal conductivity and management}
\label{sec:thermal}
Beyond efficient charge extraction, another crucial material property for real-world device application is its macroscopic thermal conductivity $\kappa$. 
In high-power electronics and concentrator photovoltaics, rapid heat dissipation is essential to prevent thermal runaway, suppress thermally activated leakage currents, and mitigate temperature-induced degradation of the electronic structure and carrier transport.
While pristine diamond is known to exhibit one of the highest thermal conductivities among three-dimensional materials, the introduction of high defect concentrations degrades heat transport due to defect-induced phonon scattering, reflected in reduced phonon lifetimes and altered dispersion.
To evaluate this performance trade-off within our proposed contact and absorber layers, we calculate the lattice thermal conductivity.
As discussed in \autoref{sec:comp_details}, the computational cost of evaluating third-order force constants for a 250-atom supercell from first principles is tremendous.
Therefore, to estimate the thermal conductivity at the device-relevant defect concentration of $0.4\%$, we employ the Matthiessen's rule assuming independent scattering processes in the dilute limit \cite{PGKlemens_1955, PhysRev.113.1046}. 
In this regime, the total thermal resistivity ($1/\kappa$) can be approximated as the sum of intrinsic phonon-phonon resistivity and defect-induced scattering contributions.
Under this approximation, phonon scattering by point defects leads to an approximately linear dependence of the lattice thermal resistivity on defect concentration:
\begin{equation}
 \frac{1}{\kappa(c)} = \frac{1}{\kappa_0} + A c
\end{equation}
where $\kappa_0$ is the thermal conductivity of pristine diamond and $A$ is a defect scattering parameter.
Using the calculated values of $\kappa_0$ and $\kappa(c)$ at $c = 6.25\%$, we extract the parameter $A$ and use it to estimate the thermal conductivity at the device-relevant concentration of $c = 0.4\%$.
We acknowledge that a $6.25\%$ concentration extends beyond the dilute limit and may introduce non-linear scattering effects.
However, we expect such non-linearities to overstate the thermal resistivity at lower concentrations \cite{PhysRevB.90.094117}.

\begin{figure}[htb]
\centering
\includegraphics[width=1.0\columnwidth]{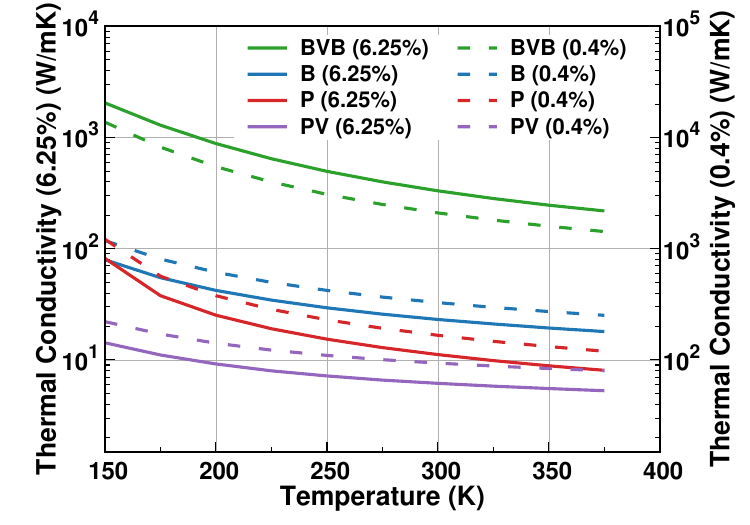}
\caption{Lattice thermal conductivity as a function of temperature for diamond with BVB, B, P, and PV defects. Solid lines correspond to explicit calculations at a defect concentration of 6.25$\%$, while dashed lines indicate extrapolated values for a device-relevant concentration of 0.4$\%$. Due to the near isotropy of thermal transport, a single Cartesian direction is shown.
}
\label{fig:thermal_conductivity}
\end{figure}

By analysing the results reported in \autoref{fig:thermal_conductivity}, it is evident that BVB-doped structure exhibits significantly higher thermal conductivity than all other systems, reaching $\sim$2058-2100 W/mK at 300 K (depending on direction) for a defect concentration of 0.4$\%$, and $\sim$315-331 W/mK for 6.25$\%$.
This is a highly favourable result, as the BVB layer constitutes the thickest region of the proposed device and is also the primary absorber of incident radiation, and therefore the dominant source of heat.
The obtained values significantly exceed the thermal conductivities of conventional photovoltaic semiconductors, such as Si ($\sim$150 W/mK), GaAs ($\sim$55 W/mK), CdTe ($\sim$10 W/mK)\cite{10.1063/5.0226632}.
Interestingly, this behaviour is consistent with the electronic transport results, indicating that the BVB defect perturbs the host lattice least among the considered defect configurations.
In contrast, the boron- and phosphorus-doped P- and N-region exhibit substantially lower thermal conductivities.
At 300 K, the boron-doped system reaches $\sim$328 W/mK at 0.4$\%$ and $\sim$23.1 W/mK at 6.25$\%$, while the phosphorus-doped system exhibits $\sim$166 W/mK and $\sim$11.2 W/mK for the same concentrations, respectively.
The lower thermal conductivity in the phosphorus-doped system is consistent with the stronger phonon scattering induced by heavier dopant atoms, as described in classical models of point-defect scattering \cite{PGKlemens_1955}.
The PV-doped system exhibits the lowest thermal conductivity among all studied structures, reaching only $\sim$88-94 W/mK at 300 K for the extrapolated concentration, and $\sim$5.8-6.2 W/mK at 6.25$\%$.
Despite the reduced thermal conductivity, the PV-doped material still exhibits values which are orders of magnitude higher than those of metal-halide perovskites (often below 1 W/mK)\cite{10.1063/5.0226632}.
Finally, it is noteworthy that the anisotropy in thermal conductivity is relatively weak compared to that observed in electronic transport properties, indicating that phonon transport remains largely isotropic despite the defect-induced symmetry breaking.

\subsection{Design of the device and limitations of the approach}
\label{sec:device}
In this section, we consolidate the design considerations derived from the previous analyses and define the corresponding device parameters.
These parameters are incorporated into a self-consistent Poisson solver to obtain the equilibrium electrostatic solution.
We analyse the resulting band bending, electric field distribution, depletion widths, and the impact of the doping profiles.

\subsubsection{Poisson solver}
The Poisson solver is formulated using the density of states (DOS) obtained from DFT calculations.
The DFT DOS is subsequently corrected by GW eigenvalue shifts, whereby the band gaps are consistently widened while preserving the underlying spectral features.
To construct a consistent band alignment across the PIN and PN junctions, we adopt an anchoring scheme as follows:
Donor-like phosphorus defects in the N-region are assumed to predominantly affect the conduction band, leaving the valence band edge largely unchanged, whereas acceptor-like boron defects in the P-region primarily modify the valence band with minimal impact on the conduction band.
The intermediate-band (I) region is treated as a reference system with unperturbed host band edges. 
Consequently, band alignment is achieved by anchoring the conduction band at the PI interface and the valence band at the IN interface. 
In the case of the PN junction, the phosphorus-vacancy defect is assumed not to significantly perturb the host band edges;
therefore, the valence band is referenced to the one of I-region, which serves only as an internal reference for alignment and is not explicitly included in the device simulation.
This approach preserves meaningful band offsets while avoiding discontinuities in the dominant carrier channels.
By energetically anchoring all spatial regions to this common reference, the built-in potential ($V_{bi}$) is consistently extracted.

Mobile n- and p-type spatial carrier concentrations are calculated via numerical integration of the GW-corrected DOS utilising Fermi-Dirac statistics.
\begin{equation}
n(x) = \frac{1}{V_{cell}} \int_{E_c}^{\infty} \text{DOS}(E)\, f(E  - e\phi(x))\, dE
\end{equation}
\begin{equation}
p(x) = \frac{1}{V_{cell}} \int_{-\infty}^{E_v} \text{DOS}(E)\, [1 - f(E - e\phi(x))]\, dE
\end{equation}
where $V_{cell}$ is the supercell volume, $\phi (x)$ is the local electrostatic potential, $E_v$ and $E_c$ are the valence and conduction band edges, respectively, and $f(E  - e\phi(x))$ is the Fermi-Dirac distribution.
For all our calculation we set the temperature to 300 K.
%
%
Intermediate band states are treated as localized charge reservoirs that are electrostatically coupled to the local potential, such that their occupation follows Fermi-Dirac statistics, while remaining spatially immobile and therefore not contributing to carrier transport.
The trapped charge density $\rho_{trap}$ is evaluated by integrating within the discrete energy bounds of the respective defect intermediate bands:
\begin{equation}
\rho_{trap}(x) = -\frac{e}{V_{cell}} \sum_{i} \int_{IB_i} \text{DOS}(E)\, f(E - e\phi(x))\, dE.
\end{equation}

The equilibrium electrostatic potential $\phi(x)$ is obtained by solving the one-dimensional nonlinear Poisson equation:
\begin{equation}
\epsilon \frac{d^2\phi}{dx^2} = -\rho_{total}(x)
\end{equation}
where the total space charge $\rho_{total}$ contains contributions from mobile carriers ($p(x)$ and $-n(x)$), ionised dopants ($N_D^+(x)$ and $-N_A^-(x)$) and localised trap states ($ \rho_{trap}(\ x)$):
\begin{equation}
    \rho_{total}(x) = e \left[ p(x) - n(x) + N_D^+(x) - N_A^-(x) \right] 
    + \rho_{trap}(x)
\end{equation}

To account for atomic interdiffusion at the junctions, the spatial profiles of dopants and defects are smoothed using hyperbolic tangent weighting functions with a characteristic transition width $\sigma$:
\begin{equation}
    N_a(x) = N_A \cdot W_P(x), \quad N_d(x) = N_D \cdot W_N(x).
\end{equation}
where, in the contact regions, $N_A$ and $N_D$ are defined by the bulk equilibrium carrier densities and the weighting functions are 
\begin{equation}
    W_P(x) = \frac{1}{2} \left[ 1 - \tanh\left(\frac{x - x_{PI}}{\sigma}\right) \right]
\end{equation}
\begin{equation}
    W_N(x) = \frac{1}{2} \left[ 1 + \tanh\left(\frac{x - x_{IN}}{\sigma}\right) \right]
\end{equation}
where $x_{PI}$ and $x_{IN}$ are the junction coordinates.
Moreover, the density of the BVB complexes is scaled by the weighting factor $W_I(x) = 1 - W_P(x) - W_N(x)$.
For the PN junction, the two interface coordinates reduce to a single junction coordinate $x_{PN}$.

Although all regions are based on diamond, their combination results in an effective heterojunction due to defect-dependent band edge shifts.
Differences in band gap values may arise both from physical effects, such as band gap narrowing at high doping concentrations, and from defect-dependent variations in GW corrections and exchange-correlation functionals.
To disentangle these contributions, the effective driving forces acting on charge carriers are decomposed into electrostatic and band-structure components.
The electrostatic field $E_{es}$ is defined as
\begin{equation}
E_{es}(x) = -\nabla \phi(x)
\end{equation}
and is governed solely by the space charge distribution. 
In contrast, electrons experience an effective quasi-electric field $E_{q}$ given by
\begin{equation}
E_q(x) = \frac{1}{e} \nabla E_c(x)
\end{equation}
which includes both electrostatic and material-dependent band-edge variations.
Here, $E_c(x)$ denotes the local conduction-band edge, which varies spatially due to the band bending caused by electrostatic potential and heterojunction-induced band offsets.
In regions with abrupt band offsets (e.g. across interfaces with $\sigma \approx 1$ nm), spatial variations in $E_c$ can produce apparent fields on the order of MV/cm. 
By comparing $E_q$ and $E_{es}$, the relative contribution of heterojunction-induced band offsets can be quantified, allowing us to distinguish between true space-charge-driven electrostatics and band-structure-induced driving forces.
Analogously to $E_q$, we can define the hole quasi-electric field $E^h_q(x) = \frac{1}{e} \nabla E_v(x)$, where $E_v(x)$ is the valence-band edge.
In the following, we focus on $E_q$, since in the configurations considered here, the hole quasi-field either coincides with the electrostatic field ($E_q^h = E_{es}$) or exhibits significantly weaker variations, particularly at the PI interface compared to the pronounced peaks of $E_q$ at the IN interface of the PIN junction (\autoref{fig:PIN_dashboard}).

\subsubsection{PIN junction for intermediate-band photovoltaics}

\begin{figure*}[htb]
\centering
\includegraphics[width=1.0\textwidth]{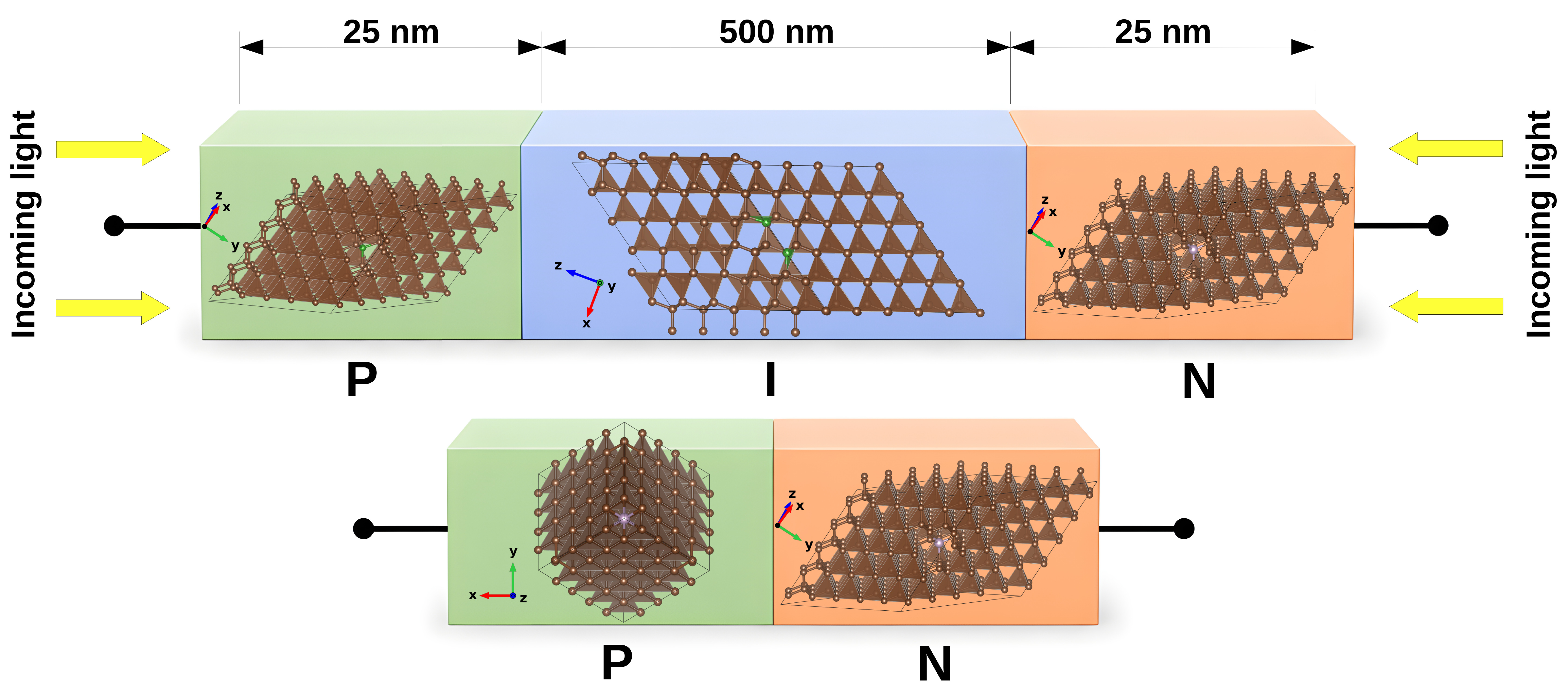}
\caption{ Schematic of the simulated diamond-based devices with explicit atomic structures.
Top: bifacial PIN junction solar cell architecture. 
The device dimensions are optimised based on modelling guidelines, featuring a 500 nm intermediate-band absorber (I-layer) bounded by highly transparent 25 nm P- and N-type contact layers.
The I-region contains boron–vacancy–boron (BVB) defects forming the intermediate bands, while the P- and N-regions are doped with substitutional boron and phosphorus, respectively.
Bottom: single-dopant-species PN junction, where the P-region is formed by phosphorus–vacancy (PV) defects and the N-region is doped with substitutional phosphorus.
In both devices, the atomic lattices, defect configurations, and crystallographic orientations correspond exactly to those used in the simulations.
Incoming light directions and electrical contacts are indicated.
}
\label{fig:device_scheme}
\end{figure*}

We begin by analysing the PIN junction device proposed for intermediate band photovoltaic cell.
We recall, that our design includes the boron-doped P-region, boron-vacancy-boron (BVB) I-region as an absorber, and phosphorus-doped N-region.  
In the previous discussion we established the following design suggestions and characteristics:
\begin{itemize}
 \item The optimal crystal orientation corresponds to incident light propagating within the \textit{xz}-plane of the simulated supercell, such that the electric field is aligned along the \textit{y}-direction, where the absorption is strongest as illustrated in \autoref{fig:device_scheme}.
 \item Increasing the absorber thickness above 500 nm does not significantly enhance absorption.
 \item Lower occupations of the intermediate bands (e.g. 0.1 electron per cell) generally yield more favourable absorption characteristics, since large occupations (e.g. one electron per cell) alter the band structure by shifting it to higher energies outside of the main peak of the solar spectrum.
 \item The ladder-like IB$_m$$\rightarrow$IB$_n$ absorption is essential for utilising low energy photons; however, efficient utilisation would require a sufficiently thick I-region.
 \item Contact P- and N-region do no block useful part of the solar spectra even when above 100 nm thick, suggesting that a bifacial device configuration may be feasible.
 \item Among the contact layers, the boron-doped P-region absorbs slightly less in the relevant energy ranges, making it a slightly better option.
 \item Relatively small changes of the carrier mobility and thermal conductivity of the absorber, compared to pure diamond, suggest that BVB defect has overall the lowest impact on the properties of the host matrix, except for adding intermediate impurity bands.
\end{itemize}

Based on these findings, we choose the width of the P and N contact layers to be 25 nm and vary the thickness of the absorber, considering 150, 500, and 1000 nm, with a doping transition parameter of $\sigma = 1.0$ nm.
Our most viable device candidate consists of an absorber with a thickness of 500 nm (\autoref{fig:device_scheme}); therefore, for this design we additionally investigate three values of $\sigma$: 0.1 nm representing an abrupt junction, and 1.0 and 3.0 nm corresponding to more diffused junctions.
Experimental studies on $\delta$-doping indicate that dopant profiles can be controlled on the nanometer scale for simple dopants \cite{C9NA00593E}.
However, we note that this behaviour may be more complex in the presence of defect complexes.

\begin{figure*}[htb]
\centering
\includegraphics[width=1.0\textwidth]{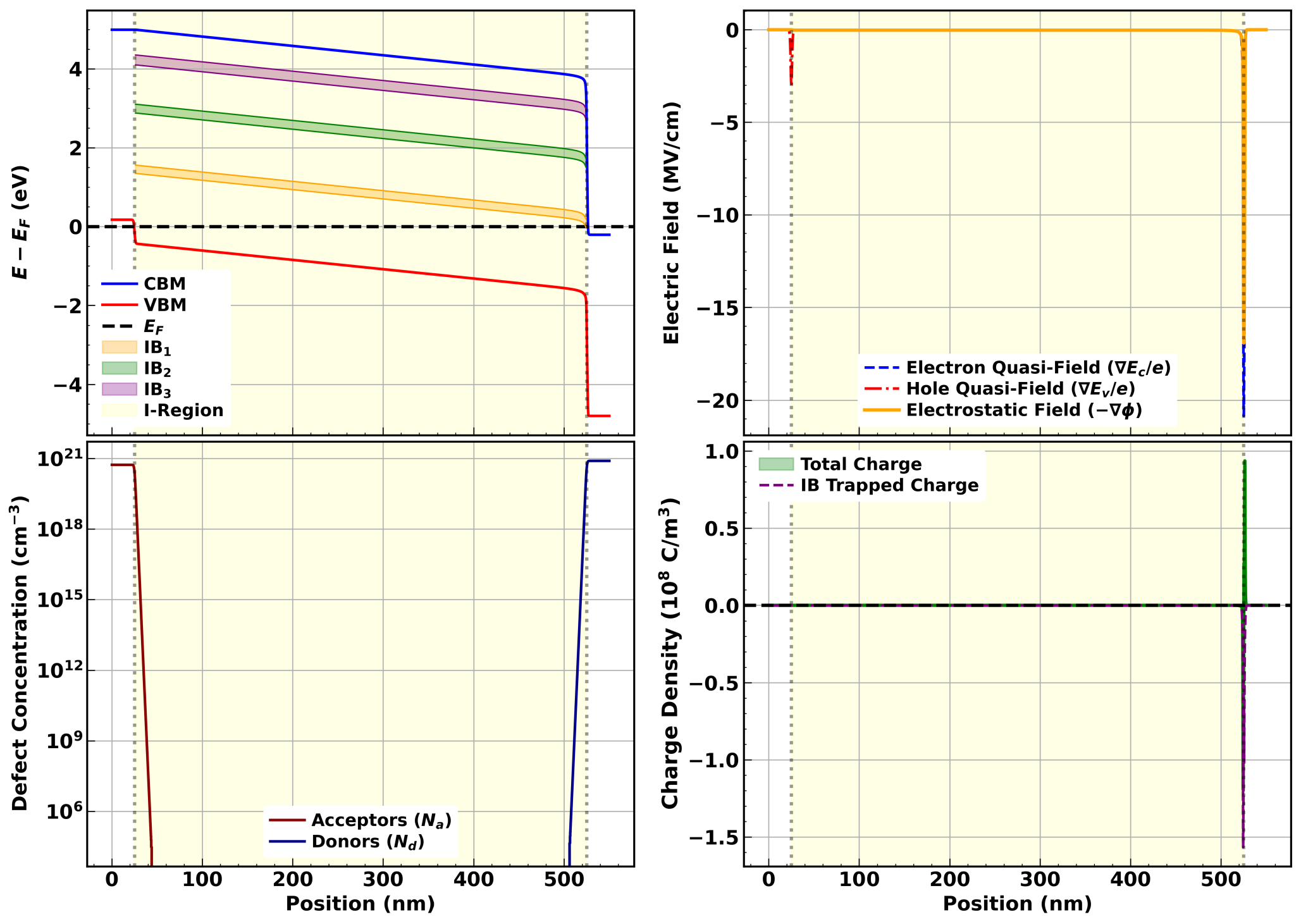}
\caption{Equilibrium PIN junction in diamond with degenerately doped boron (P-region) and phosphorus (N-region), and a 500 nm thick absorbing BVB-doped I-layer represented by a yellow overlay. Panels show: \textit{a)} band alignment including intermediate bands, \textit{b)} internal electric field profile highlighting the IN junction spike, \textit{c)} logarithmic doping profiles of the P- and N-regions for $\sigma = 1.0$ nm, and \textit{d)} net space charge density showing the depletion region dipole. In the electric field profile, the electrostatic field ($E_{es}$) and quasi-fields overlap in most regions. Deviations occur at the interfaces: a hole quasi-field peak appears at the PI interface ($E_q^h > E_q = E_{es}$), while a more pronounced electron quasi-field peak is observed at the IN interface ($E_q > E_q^h = E_{es}$).
}
\label{fig:PIN_dashboard}
\end{figure*}

The first observation from the analysis of \autoref{fig:PIN_dashboard} is that the pronounced charge and electric field spikes are localised exclusively at the IN junction.
This indicates that the I-region behaves effectively as a weakly-doped p-type region, which directly follows from the position of the Fermi level located between the valence band and IB$_1$ as seen in \autoref{fig:BS}.
Another significant result is the reduction of the electric field within the I-region ($\sim$0.024 MV/cm) compared to a PIN junction with a pristine diamond ($\sim$0.086 MV/cm) for a 500 nm device, as shown in \autoref{tab:PIN}.
This behaviour originates from the presence of intermediate-band states, which impose a direct constraint on the band bending within the I region.
In particular, the lowest intermediate band (IB$_1$), located above the Fermi level, limits the steepness of the potential gradient (\autoref{fig:PIN_dashboard}).
A steeper band bending would force IB$_1$ below the Fermi level deeper within the bulk, leading to its occupation and the buildup of localised charge.
This additional charge modifies the space-charge distribution and, through the Poisson equation, counteracts further increase of the electric field.
As a result, the system limits the potential gradient, keeping IB$_1$ above the Fermi level throughout most of the I-region and allowing it to cross the Fermi level only near the IN junction. 
We refer the reader to also inspect the band bending, space charge and electrostatic field profiles of the pristine diamond PIN junction in the Supplementary Material Section II for comparison with intermediate band PIN junction.

\begin{figure}[htb]
\centering
\includegraphics[width=1.0\columnwidth]{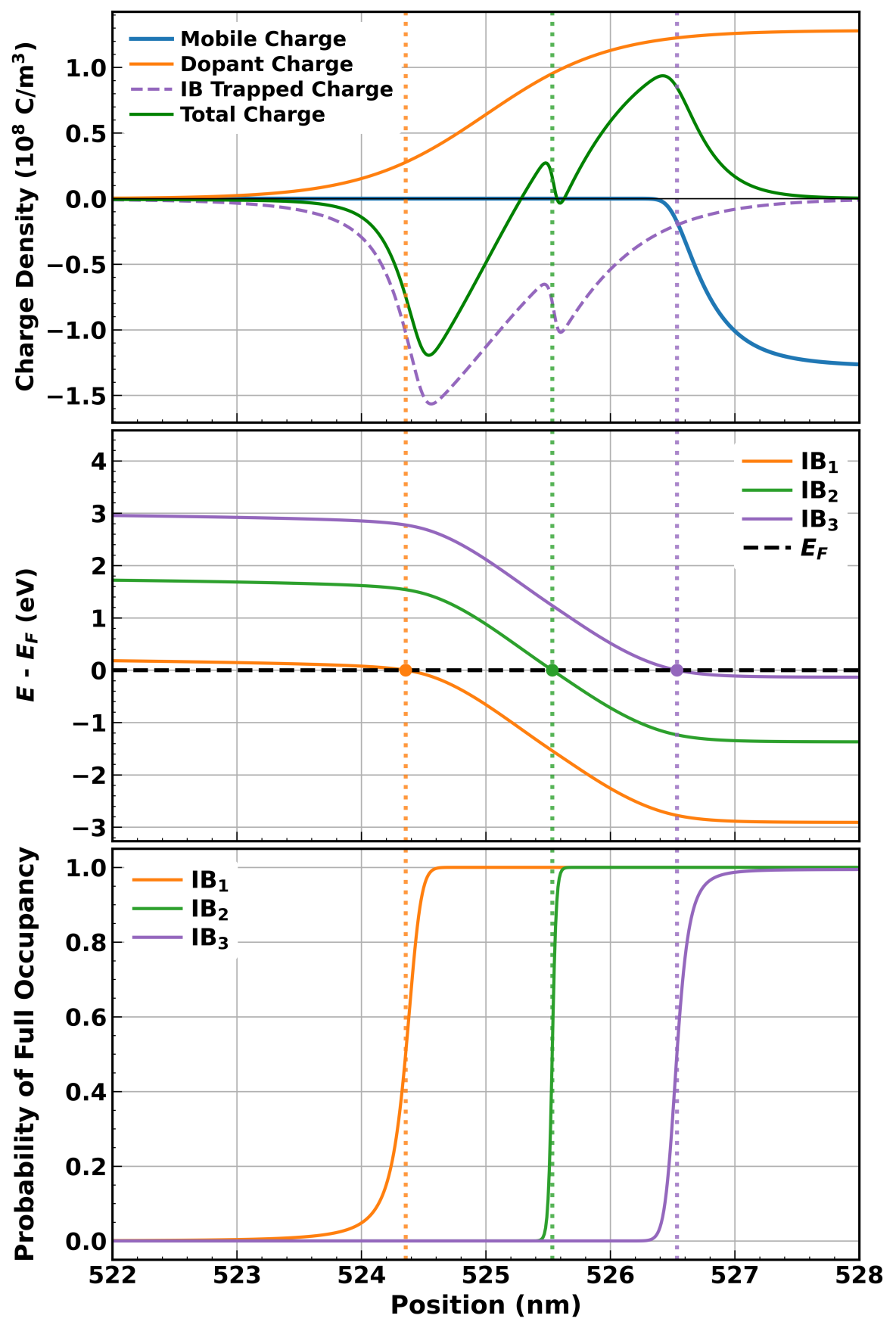}
\caption{Detailed analysis of the IN junction interface with $\sigma = 1.0$ nm: \textit{a)} spatial charge decomposition near the IN interface, \textit{b)} energy levels of the centre of the intermediate bands relative to Fermi level $E_F$, identifying the exact locations of $E_F$ crossing, and \textit{c)} intermediate band full occupancy probability illustrating the trap filling that generates the localised charge spikes seen in Panel (a). 
}
\label{fig:IN_charge}
\end{figure}

Moreover, by analysing a detail of the spatial charge distribution (\autoref{fig:IN_charge}a) around the IN interface, we observe significant non-monotonic features.
In a conventional junction, we would expect a single negative charge accumulation on the P-side (or I-side in this case) and a corresponding positive charge accumulation on the N-side, resulting in two extrema.
However, in the intermediate-band PIN device, multiple extrema are present.
This behaviour arises from the sequential filling of individual intermediate bands, as each band undergoes a sharp occupation transition when shifted below the Fermi level (\autoref{fig:IN_charge}a, b), leading to variations in trapped charge.
Furthermore, each BVB defect can accommodate up to six electrons, whereas a phosphorus dopant contributes with only one.
This combined with the gradual spatial doping profiles, causes the onset of mobile charge to occur relatively far from the junction. 

We proceed by analysing the effects of doping profiles and absorber thicknesses.
The built-in potential remains constant across all configurations due to the consistent DFT-aligned Fermi level difference between the P- and N-region.
Therefore, as results in \autoref{tab:PIN} show, the depletion width scales directly with the I-region thickness, confirming that the absorber is fully depleted in all cases.
The depletion region is defined as a region where the  electric field exceeds 0.03 $\%$ of the peak junction field.   
Correspondingly, the electric field within the I-region decreases approximately inversely with its width, as expected from electrostatic considerations.
Furthermore, the absorber region thickness does not influence the peak fields or peak charge densities.

In contrast, we observe a strong dependence of the peak electric field and charge density on the junction smoothing parameter $\sigma$. 
For an abrupt junction ($\sigma = 0.1$ nm), extremely large quasi-electric field ($E_q$) is observed, significantly exceeding the electrostatic field ($E_{es}$).
This originates from steep spatial variations of the conduction band edge, reflecting heterojunction-induced band offsets rather than space-charge effects. 
As $\sigma$ increases, both fields are progressively suppressed, with quasi-electric fields decreasing more rapidly leading to convergence of $E_q$ towards $E_{es}$ and a reduction of peak charge densities. 
Moreover, the electrostatic field within the intrinsic region remains identical for both $E_{es}$ and $E_q$ across all configurations, confirming that carrier separation in the absorber is governed by the macroscopic space-charge distribution.
The discrepancy between $E_q$ and $E_{es}$ is therefore localised to the junction regions and serves as a quantitative measure of heterojunction-induced band-edge effects.
These results highlight the importance of carefully treating band alignment as it may introduce artificial fields related to the DFT or GW band offsets.
Furthermore, it is important to note that steep doping profiles, resulting in high electric field may lead to unwanted IB$_m$$\rightarrow$CB tunneling at the IN interface in the real device, potentially degrading its performance.
This motivates the use of graded junctions to preserve the effective isolation of the intermediate bands.

From a photovoltaic device perspective, the bulk electric field within the absorber is critical for efficient carrier extraction before recombination processes can occur.
As observed, the bulk field decreases with increasing absorber thickness, creating a trade-off between enhanced optical absorption and reduced electric field.
However, due to the degenerately doped P- and N-regions, the electric field remains high even at 1000 nm ($\sim$11.4 kV/cm).
Combined with the high carrier mobility of the BVB layer ($\sim 1000$-$2000$ cm$^2$/Vs) at 300 K, depending on carrier type and direction), this yields drift velocities ($v_d = \mu E$) on the order of $10^7$ cm/s.
These values approach the saturation velocity in diamond, which lies in the range of $0.85$-$1.2\times10^7$ cm/s for holes and $1.5$-$2.7\times10^7$ cm/s for electrons \cite{WORT200822}.
Therefore, even at 1000 nm absorber thickness, the electric field remains sufficient for efficient carrier extraction.

\begin{table*}[htbp]
\centering
\begin{tabular}{c|c|c|c|c|c|c|c|c|c}
\hline
\textbf{Device} & \textbf{$\sigma$} & \textbf{$W$} & \multicolumn{2}{c|}{\textbf{Peak Charge ($10^8$ C/m$^3$)}} & \multicolumn{3}{c|}{\textbf{Peak Field (MV/cm)}} & \multicolumn{2}{c}{\textbf{Bulk Field (MV/cm)}} \\
\hline
(P-I-N nm) & (nm) & (nm) & P-Side & N-Side & $E_{es}$ & $E_q$ & $\Delta E$ & $E_{es}$ & $E_q$ \\
\hline
25-150-25 & 1.0 & 154.00 & -1.19 & 0.936 & 16.90 & 20.85 & 3.95 & 0.0843 & 0.0843 \\
\hline
25-500-25 & 0.1 & 502.15 & -2.26 & 1.280 & 28.21 & 69.96 & 41.75 & 0.0235 & 0.0235 \\
25-500-25 & 1.0 & 503.40 & -1.19 & 0.936 & 16.90 & 20.85 & 3.95 & 0.0237 & 0.0237 \\
25-500-25 & 3.0 & 506.92 & -0.548 & 0.491 & 10.37 & 11.66 & 1.29 & 0.0243 & 0.0243 \\
\hline
25-1000-25 & 1.0 & 1002.78 & -1.19 & 0.936 & 16.90 & 20.85 & 3.95 & 0.0114 & 0.0114 \\
\hline
\hline
\textbf{25-500-25*} & 1.0 & 505.06 & -0.002 & 0.002 & 0.09 & 4.63 & 4.54 & 0.0860 & 0.0860 \\
\hline
\multicolumn{10}{l}{\small *Pristine diamond I-region.}
\end{tabular}
\caption{Numerical analysis of space charge and field characteristics across varying absorption layer thicknesses and doping transition widths ($\sigma$). The built-in potential is constant ($V_{bi} = 4.3639$ V) for all cases, as determined by the DFT-aligned Fermi levels of the P- and N-region. $E_{es}$, $E_q$ and $\Delta E$ correspond to electrostatic field, electron quasi-field and their difference, respectively. The final row represents the pristine (pure diamond) control case.}
\label{tab:PIN}
\end{table*}

\subsubsection{PN junction based on phosphorus defects}
Our second studied architecture is a PN junction formed solely from phosphorus-based defects.
The P-region is doped with phosphorus-vacancy (PV) complexes, while the N-region is doped with substitutional phosphorus.
In the previous sections, we established the following design considerations and material characteristics:
\begin{itemize}
 \item Promoting an electron to the impurity band while creating a hole in the valence band requires approximately 0.67 eV. 
 Consequently, at 300 K the concentration of thermally activated carriers in the valence band is negligible.
 \item The thickness of the impurity band combined with relatively high electronic conductivity and low carrier mobility, indicates that impurity-band conduction is the dominant transport mechanism.
 This makes a PV-doped diamond a promising degenerate p-type semiconductor for applications such as tunnel diodes, PIN junctions, or asymetrically-doped (P$^+$N) junctions, while its use in a conventional rectifying diodes is significantly limited.
 \item The conductivity and carrier mobility of PV-doped diamond are approximately a factor of two higher in \textit{xy}-plane than along the \textit{z}-axis.
 High conductivity and mobility are particularly advantageous for reducing power losses and enabling fast switching in high-power and high-frequency electronics \cite{WORT200822}.
 Therefore, the preferred crystal orientation for the PV-defect system is such that the \textit{z}-direction of the simulated supercell is aligned parallel to the junction plane, as illustrated in \autoref{fig:device_scheme}.
 \item The Seebeck coefficient of PV-doped diamond exhibits anisotropic and bipolar transport behaviour within the impurity band.
 At low temperatures, the Seebeck coefficient is negative across all Cartesian directions, indicating electron-dominated transport, whereas at higher temperatures thermal excitation leads to a transition towards hole-dominated transport. 
 This may be utilised for mitigation of parasitic thermal voltages or for thermal management applications. 
\end{itemize}

\begin{figure*}[htb]
\centering
\includegraphics[width=1.0\textwidth]{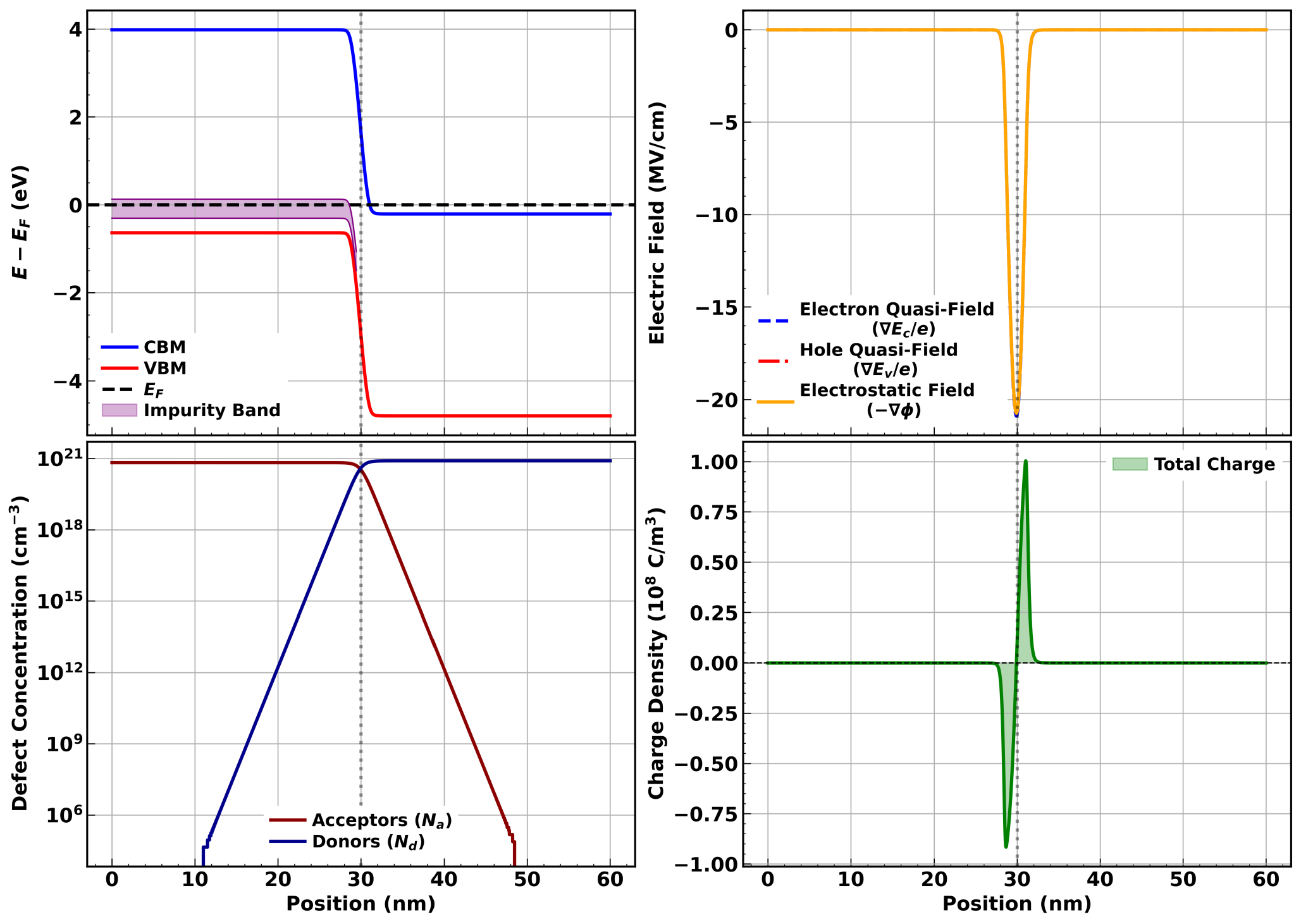}
\caption{Equilibrium PN junction in diamond with a PV-doped P-region and single substitutional phosphorus-doped N-region.
Panels show: \textit{a)} band alignment, \textit{b)} internal electric field profile highlighting the junction spike, \textit{c)} logarithmic doping profiles of the P- and N-regions for $\sigma = 1.0$ nm, and \textit{d)} net space charge density illustrating the depletion-region dipole.
In the electric field profile, the electrostatic field ($E_{es}$) and quasi-fields overlap in most regions. A slight deviation occurs only at the PN interface, where the electron quasi-field exhibits a marginally larger amplitude ($E_q > E_q^h = E_{es}$).
}
\label{fig:PN_dashboard}
\end{figure*}

Since the junction geometry is not a primary variable in this case, we fix the layer thicknesses to 30 nm for both P- and N-region and analyse only the effect of the doping transition parameter $\sigma$.
This thickness is sufficient to fully accommodate the depletion region and ensure that the electrostatic potential and carrier densities reach their bulk-like behaviour away from the interface.
Similarly to the PIN junction, we consider $\sigma = 0.1$ nm as an abrupt junction, and $\sigma = 1.0$ nm and $3.0$ nm as progressively more diffused profiles.
It is important to note that, although the Seebeck coefficient indicates anisotropic and temperature-dependent bipolar transport within the impurity band, these effects do not influence the equilibrium electrostatics considered here. 
The Poisson solver depends solely on the occupation of electronic states, which is fully captured by the density of states and Fermi-Dirac statistics.
Since transport in the PV-doped region is predominantly mediated by the impurity band, the solver treats this band as an effective extension of the valence band, in contrast to the intermediate bands in the PIN absorber, which are treated as trap states.
Furthermore, the PV-induced impurity band is partially filled, corresponding approximately to one hole per defect state.
This mirrors the contribution of substitutional phosphorus in the N-region, which provides approximately one free electron per dopant.
As a result, the effective acceptor and donor concentrations are comparable, leading to a relatively symmetric space-charge distribution across the junction (\autoref{fig:PN_dashboard}).

As expected, the results presented in \autoref{tab:PN_junction} show that both the electrostatic field $E_{es}$ and the quasi-electron field $E_q$ decrease with increasing doping transition width $\sigma$.
Similarly to the PIN junction, the quasi-electric field decreases more rapidly, leading to convergence of $E_q$ towards $E_{es}$ and a reduction of peak charge densities.
In contrast to the PIN case, however, the discrepancy between $E_{es}$ and $E_q$ is significantly smaller.
This indicates that heterojunction-induced effects are strongly suppressed in a junction formed between two degenerately doped regions.
This behaviour originates from the reduced band-edge mismatch between the two sides of the junction: in the PV-doped region, transport occurs within a partially filled impurity band acting as an extension of the valence band, while in the N-region the conduction band is similarly shifted due to degenerate doping.
As a result, the effective band offsets at the junction are much smaller than in the PIN case, where the BVB-doped region retains the wide band gap of diamond, leading to pronounced band discontinuities at the interfaces.

\begin{table}[htbp]
\centering
\setlength{\tabcolsep}{3pt} 
\begin{tabular}{c|c|c|c|c|c|c|c}
\textbf{Device} & \textbf{$\sigma$} & \textbf{$W$} & \multicolumn{2}{c|}{\textbf{\shortstack{Peak Charge \\ ($10^8$ C/m$^3$)}}} & \multicolumn{3}{c}{\textbf{\shortstack{Peak Field \\ (MV/cm)}}} \\
\hline
(P-N nm) & (nm) & (nm) & P-Side & N-Side & $E_{es}$ & $E_q$ & $\Delta E$ \\
\hline
30-30 & 0.1 & 4.81 & -1.080 & 1.290 & 29.09 & 30.58 & 1.49 \\
30-30 & 1.0 & 6.23 & -0.916 & 1.000 & 20.74 & 20.88 & 0.14 \\
30-30& 3.0 & 15.44 & -0.558 & 0.575 & 14.94 & 14.99 & 0.05 \\
\hline
\end{tabular}
\caption{Numerical analysis of space charge and field characteristics for the PN junction across varying doping transition widths ($\sigma$). The built-in potential is constant ($V_{bi} = 4.1623$ V) for all cases, as determined by the DFT-aligned Fermi levels of the P- and N-region. $E_{es}$, $E_q$, and $\Delta E$ represent the electrostatic field, electron quasi-field, and the heterojunction-induced field contribution, respectively.}\label{tab:PN_junction}
\end{table}

\subsubsection{Limitations of the approach}
The present study combines analysis of the first-principles electronic structure, optical properties, electron-phonon coupling, and electrostatic device modelling to provide a consistent picture of the proposed architectures.
However, several limitations should be noted.

Firstly, the transport properties are evaluated within the phonon-limited regime, neglecting additional scattering mechanisms such as ionised impurity scattering \cite{PhysRevMaterials.6.L010801, PhysRevB.107.125207}, or interface scatterings. 
As a result, the calculated mobilities and conductivities represent upper bounds on carrier transport, particularly in the degenerately doped regions where Coulomb scattering is expected to be significant.

Secondly, the thermal transport is estimated using Matthiessen’s rule and extrapolation from high defect concentrations, which assumes approximately independent scattering channels and linear scaling of resistivity with defect concentration.
While this provides a reasonable estimate of trends, deviations from the dilute limit may be present.

Finally, a full quantitative assessment of device performance would require explicit modelling of carrier recombination processes, particularly Shockley-Read-Hall recombination mediated by intermediate-band states.
Such processes are known to play a critical role in intermediate-band photovoltaics, where mid-gap states can act as efficient recombination centres.
In this context, the evaluation of carrier lifetimes and generation rates would enable the solution of non-equilibrium continuity equations and the extraction of current-voltage characteristics.

Despite these limitations, the present approach provides a  first-principles-based framework for assessing the interplay between electronic structure, optical properties, transport and device-level electrostatics, and establishes physically grounded design guidelines for diamond-based photovoltaic and diode architectures.

\section{Conclusions}
\label{sec:conclusions}

In this work, we assess diamond-based photovoltaic and diode architectures within a first-principles framework.  
We calculate the electronic structure using density functional theory with GW correction, optical properties via the Bethe-Salpeter equation and independent-particle approximation, carrier transport with electro-phonon coupling, and evaluate lattice thermal conductivity features.
These results are subsequently used as input for a Poisson solver, enabling a description of device-level electrostatics. 
This approach provides a unified methodology for linking electronic structure, optical response, transport and junction behaviour.
We investigate two device architectures: a PIN junction composed of boron-doped (P), boron-vacancy-boron (BVB) intermediate-band (I), and phosphorus-doped (N) regions, and a PN junction formed entirely from phosphorus-based defects, namely phosphorus-vacancy (PV) and substitutional phosphorus.
We first establish key theoretical aspects, including the importance of GW corrections for accurate band gap widths and impurity levels, as well as the energy dependence of excitonic shifts in the BVB absorber, before analysing the results from a device-oriented perspective.

For the PIN architecture, the absorber's BVB defect introduces multiple intermediate bands while preserving the favourable transport and thermal properties of the diamond host. 
The optical absorption exhibits anisotropy, with optimal overlap with the solar spectrum achieved when the electric field of incident light is aligned along the direction of strongest absorption, corresponding to light propagation in the \textit{xz}-plane.
The analysis of excited-state absorption shows that moderate intermediate-band occupations enhance absorption across the solar spectrum, whereas high occupations significantly distort the band structure and reduce spectral overlap with the solar spectrum.
Furthermore, the contact P- and N-region do not block a significant portion of the useful solar spectrum even at thicknesses above 100 nm, suggesting that bifacial device configurations may be feasible.
Among the contact layers, the boron-doped P-region exhibits slightly lower absorption in the relevant energy range, making it a preferable option.
The electronic conductivity and carrier mobility of both the contact and absorber regions are sufficiently high to support efficient carrier extraction.
From a device perspective, an absorber thickness of approximately 500 nm provides an optimal balance between absorption, electric field strength, and material usage.
Although the electric field decreases with increasing absorber thickness, it remains sufficiently high to sustain carrier drift velocities approaching the saturation limit of diamond, ensuring efficient carrier extraction even at 1000 nm.
The electrostatic analysis further reveals that the presence of intermediate bands imposes a constraint on band bending within the absorber, effectively limiting the internal electric field. 
Moreover, abrupt junctions lead to strong electric fields that may promote unwanted tunnelling processes, highlighting the importance of graded junctions.

For the PN architecture based on phosphorus defects, transport in the PV-doped structure is governed by impurity-band conduction, as thermal activation from the valence band is negligible at room temperature. 
The PV-induced impurity band behaves as an effective extension of the valence band, leading to degenerate transport with relatively high conductivity despite reduced carrier mobility.
This makes a PV-doped diamond a promising degenerate p-type semiconductor for applications such as tunnel diodes, PIN junctions, or asymetrically-doped (P$^+$N) junctions, while its use in a conventional rectifying diodes is limited.
Additionally, the Seebeck coefficient exhibits anisotropic and sign-changing behaviour, transitioning from negative (electron-dominated) values at low temperatures to positive (hole-dominated) values at higher temperatures, which may offer opportunities for mitigating parasitic thermoelectric effects or enabling novel thermal management strategies.

Overall, our results demonstrate that the realisation of both intermediate-band photovoltaic absorbers and unconventional junction architectures in diamond is a promising avenue for further study.  
The BVB defect emerges as a particularly promising candidate, combining strong optical functionality with minimal degradation of transport and thermal properties. 
At the same time, the PV-based architecture highlights the potential of impurity-band transport for simplified, single-dopant device designs.
Moreover, even when defect incorporation reduces the electronic mobility and thermal conductivity relative to pristine diamond, the resulting values still exceed those of most conventional semiconductor materials.
These findings establish key design principles for diamond-based devices and provide a foundation for future studies incorporating non-equilibrium carrier dynamics and recombination processes.

\section*{Acknowledgements}
This work was supported by the project ``The Energy Conversion and Storage'', funded as project No. CZ.02.01.01/00/22\_008/0004617 by Programme Johannes Amos Commenius, call Excellent Research.
This work was supported by the Ministry of Education, Youth and Sports of the Czech Republic through the e-INFRA CZ (ID:90254).
The access to the computational infrastructure of the OP VVV funded project CZ.02.1.01/0.0/0.0/16\_019/0000765 ``Research Center for Informatics'' is also gratefully acknowledged.

\bibliographystyle{elsarticle-num}
\biboptions{sort&compress}

\clearpage
\includepdf[pages=-]{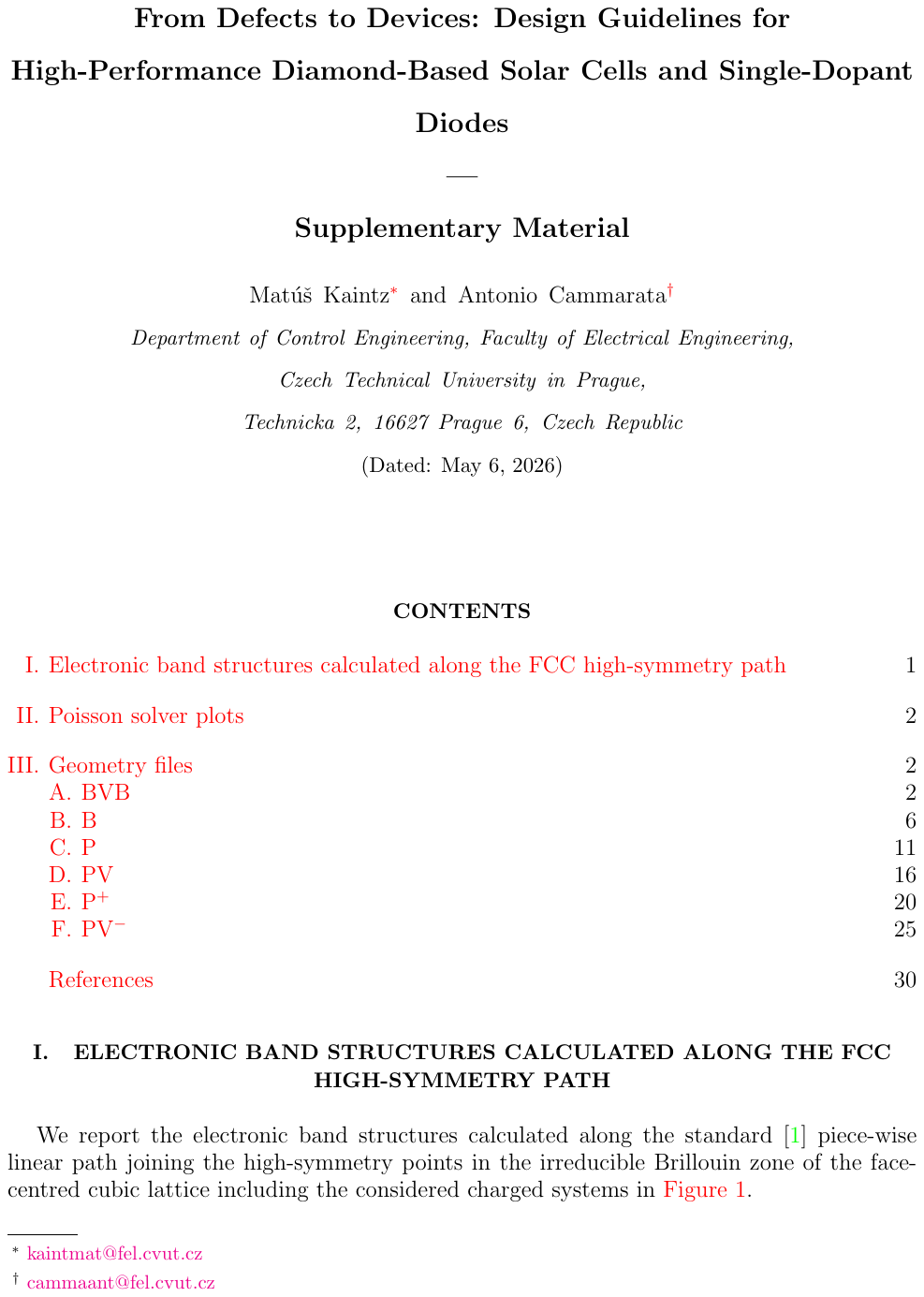}

\end{document}